\documentclass[11pt,british]{article}
\usepackage[utf8]{inputenc}
\usepackage{mathtools}
\usepackage{amsmath}
\usepackage{amssymb}
\usepackage{setspace}
\usepackage{microtype}
\makeatletter
\usepackage{tikz}
\usetikzlibrary{backgrounds}
\usepackage[framemethod=tikz]{mdframed}
\AtBeginEnvironment{mdframed}{%
\tikzset{every picture/.style={}}%
}
\mdfsetup{roundcorner=.5ex}
{\begin{mdframed}[backgroundcolor=black!10!white]}%
{\end{mdframed}}

\usepackage{jheppub}                   

\makeatletter
\gdef\@fpheader{\ }                    
\makeatother

\usepackage{setspace}
\usepackage{slashed}	 	
\usepackage{amsfonts} 		
\SetTracking{encoding={*}, shape=sc}{0} 	
\usepackage{color} 		
\definecolor{darkblue}{rgb}{0.0,0.0,0.3} 	
\allowdisplaybreaks		
\date{\today} 		
\numberwithin{equation}{section}	

\makeatletter
\g@addto@macro\bfseries{\boldmath}
\makeatother

\DeclareMathOperator{\vol}{vol}					
\DeclareMathOperator{\tr}{tr}					
\DeclareMathOperator{\re}{Re}					
\DeclareMathOperator{\im}{Im}					
\DeclareMathOperator{\ad}{ad}					
\newcommand{\AdS}[1]{\text{AdS}_{#1}}			

\newcommand{\quotient}{/}							
\newcommand{\qquotient}{/\!\!/}				
\newcommand{\eqspace}{\mathrel{\phantom{=}}{}} 				
\newcommand{\ext}{\mbox{\large $\wedge$}} 					
\newcommand{\dd}{\mathrm{d}} 								
\newcommand{\ee}{\mathrm{e}} 								
\newcommand{\ii}{\mathrm{i}} 								
\newcommand{\Dorf}{L}       

\newcommand{\rep}[1]{\boldsymbol{#1}} 						

\newcommand\qqq{\qquad\quad} 								

\newcommand{\GL}[1]{\mathrm{GL}(#1)}
\newcommand{\SL}[1]{\mathrm{SL}(#1)}
\newcommand{\SU}[1]{\mathrm{SU}(#1)}
\newcommand{\Uni}[1]{\mathrm{U}(#1)}

\newcommand{\Spinstar}[1]{\mathrm{Spin}^{*}(#1)}

\newcommand{\Orth}[1]{\mathrm{O}(#1)}
\newcommand{\SO}[1]{\mathrm{SO}(#1)}

\newcommand{\Ex}[1]{\mathrm{E}_{#1}}

\newcommand{\Gx}[1]{\mathrm{G}_{#1}}

\newcommand{\ex}[1]{\mathfrak{e}_{#1}}

\newcommand{\gl}[1]{\mathfrak{gl}_{#1}}

\title{Marginal deformations of 3d $\mathcal{N}=2$ CFTs from AdS$_4$ backgrounds in generalised geometry}

\author[a,b]{Anthony Ashmore}
\emailAdd{ashmore@maths.ox.ac.uk}

\affiliation[a]{Merton College, University of Oxford,\\ Merton Street, Oxford, OX1 4JD, UK}
\affiliation[b]{Mathematical Institute, University of Oxford, Andrew Wiles Building,\\ Woodstock Road, Oxford, OX2 6GG, UK}
\abstract{We study exactly marginal deformations of 3d $\mathcal{N}=2$ CFTs dual to $\AdS4$ solutions in eleven-dimensional supergravity using generalised geometry. Focussing on Sasaki–Einstein backgrounds, we find that marginal deformations correspond to turning on a four-form flux on the internal space at first order. Viewing this as the deformation of a generalised structure, we derive a general expression for the four-form flux in terms of a holomorphic function. We discuss the explicit examples of S$^{7}$, Q$^{1,1,1}$ and M$^{1,1,1}$ and, using an obstruction analysis, find the conditions for the first-order deformations to extend all orders, thus identifying which marginal deformations are exactly marginal. We also show how the all-orders $\gamma$-deformation of Lunin and Maldacena can be encoded as a tri-vector deformation in generalised geometry and outline how to recover the supergravity solution from the generalised metric.}

\makeatother

\usepackage{babel}
\begin{document}
\maketitle

\begin{center}
\par\end{center}

\section{Introduction}

The AdS/CFT correspondence relates gravitational theories on anti-de Sitter space to a conformal field theory living its boundary~\cite{Maldacena99}. This dual picture allows us to understand one theory in a complicated limit, such as strong coupling, by computing quantities in a computationally feasible limit in the other theory.

Conformal field theories are fixed points of the renormalisation group flow. This means the beta functions for the couplings of the theory all vanish so that the couplings do not flow. Generically, a theory will have as many beta functions as couplings, so that the conformal symmetry is present only for fixed values of the couplings – CFTs are usually isolated points in the space of couplings. CFTs with supersymmetry are somewhat special – supersymmetry leads to non-renormalisation theorems which reduce the number of independent beta functions. In many cases, these theorems constrain the beta functions to be linear combinations of the anomalous dimensions of the fields. This means that, unlike their non-supersymmetric counterparts, superconformal field theories (SCFTs) are generically part of a family of conformal theories connected by varying values of the couplings. This set of couplings describes a manifold in the space of couplings, known as the conformal manifold $\mathcal{M}_{\text{\text{c}}}$. Operators that preserve conformality classically are known as marginal. If the operators also preserve the symmetry at the quantum level – after loop corrections – they are known as exactly marginal. It is these exactly marginal couplings that describe the conformal manifold, where the number of these couplings gives the dimension of $\mathcal{M}_{\text{c}}$.\footnote{It is possible to examine the conformal manifold using conformal perturbation theory and so derive constraints that even non-supersymmetric theories should satisfy to admit exactly marginal deformations \cite{Bashmakov:2017rko}. Despite this, all currently known theories with conformal manifolds are supersymmetric. See \cite{Cordova:2016xhm} for a recent classification.} This paper will focus on the supergravity realisation of marginal deformations in 3d $\mathcal{N}=2$ theories.

Much like 4d $\mathcal{N}=1$ theories, 3d $\mathcal{N}=2$ theories admit conformal manifolds that are K\"ahler~\cite{Strassler:1998iz,Tachikawa06,Gaiotto:2007qi,MS08,Asnin:2009xx,Akerblom:2009gx,BPS10,BPS10b,CY10,BP11,ALMTW14} where the metric is the Zamolodchikov metric built from two-point functions of the exactly marginal operators~\cite{Zamolodchikov:1986gt}. The dimension of the conformal manifold can be determined either by a Leigh–Strassler type argument~\cite{LS95} or from general properties of supersymmetric field theories~\cite{Kol02,Kol10,GKSTW10}. The second approach can also be applied to strongly coupled theories without Lagrangian descriptions. Locally the conformal manifold is given by the space of marginal couplings quotiented by the complexification of the broken global symmetry group, equivalent to a symplectic quotient where the moment maps are the $D$-term constraints. This gives a local description of the conformal manifold provided one knows the operators and global symmetry of the theory. A general method for understanding the global structure of $\mathcal{M}_{\text{c}}$ is not known, though it is possible to make progress in specific cases by considering discrete symmetries and IR dualities~\cite{Baggio:2017mas}.

Many SCFTs have dual descriptions in terms of AdS solutions of string or M-theory. For example, a large class of SCFTs are dual to branes placed in conical Calabi–Yau geometries. The existence of exactly marginal operators in a SCFT then implies there is family of dual AdS solutions, specified by different metrics, choices of flux, etc. If we want to learn about the conformal manifolds we might want to start by better understanding the moduli space of supersymmetric AdS flux vacua.

Understanding the solutions that fill out this moduli space is a difficult problem. In the simplest case of type IIB on $\text{AdS}^{5}\times\text{S}^{5}$, perturbative calculations identified supersymmetric three-form flux perturbations, dual to marginal deformations of the field theory~\cite{AKY02,GP01}. An obstruction was found at third order in the deformation, reminiscent of the one-loop beta-function. Later, Lunin and Maldacena proposed a simple method for generating new AdS solutions from backgrounds possessing at least two $\Uni{1}$ isometries~\cite{LM05}. The new solutions are dual to CFTs deformed by a special class of exactly marginal deformations. For $\text{AdS}_{5}\times\text{S}^{5}$, the new solutions are dual to the exactly marginal deformation of $\mathcal{N}=4$ super-Yang–Mills known as the $\beta$-deformation. The technique can also be applied to other $\text{AdS}_{5}$ backgrounds in type IIB, in particular those with a quiver gauge theory dual~\cite{BH05,BFMMPZ08}. Unlike the earlier perturbative approach, the solution-generating technique gives the supergravity backgrounds to all orders in the deformation. Four-dimensional SCFTs generically admit deformations other than the $\beta$-deformation, but little progress has been made in constructing the full dual backgrounds.

The solution-generating technique of Lunin and Maldacena also applies to M-theory backgrounds with three $\Uni{1}$ isometries, where it has been used to find new AdS$_{4}$ solutions starting from $\text{S}^{7}$, $\text{M}^{1,1,1}$, $\text{Q}^{1,1,1}$ and others~\cite{LM05,GLMW05,AV05}. Unlike $\text{AdS}^{5}\times\text{S}^{5}$, there has not been a perturbative analysis of the marginal deformations of $\text{AdS}_{4}\times\text{S}^{7}$, however there is some guidance from the dual field theory. The gravity solution preserves $\mathcal{N}=8$ supersymmetry, or 32 supercharges, and arises as the near-horizon limit of a stack of $N$ M2-branes in flat space. The dual three-dimensional SCFT living on the branes has an $\SO{8}$ global symmetry coming from the eight directions transverse to the branes. The dual theory is expected to be strongly coupled and does not have a known Lagrangian description. Taking a $\mathbb{Z}_{k}$ quotient of $\text{S}^{7}$ leads to an $\AdS4$ solution in type IIA that is dual to ABJM which manifests only 3/4 of the supersymmetry and is weakly curved for $k\ll N\ll k^{5}$~\cite{ABJM08}. 

Although the full $\mathcal{N}=8$ theory is not known, there has been a proposal for the number of exactly marginal deformations~\cite{Kol02}. The couplings that preserve $\mathcal{N}=2$ define a conformal manifold which is locally given by
\begin{equation}
\mathcal{M}_{\text{c}}=\rep{35}\quotient\SL{4;\mathbb{C}}=\rep{35}\qquotient\SU{4},
\end{equation}
where $\SU{4}$ is the global symmetry group broken by the couplings, and $\rep{35}$ is the rank-four symmetric tensor of $\SU{4}$. We expect the exactly marginal deformations to be controlled by 20 complex parameters, in agreement with the results of Green et al.~\cite{GKSTW10}. One goal of this paper is to verify this result directly in supergravity.

Here we will study families of $\mathcal{N}=2$ $\AdS4$ backgrounds in eleven-dimensional supergravity using generalised geometry~\cite{CSW11,CSW14,PW08,GLSW09}. We will do this by starting from $\text{AdS}_{4}\times M$ Freund–Rubin type solutions, where $M$ is Sasaki–Einstein, and turning on a four-form flux perturbation that preserves supersymmetry. We will find the general form of the four-form flux to first order in the deformation parameter and identify which deformations survive to all orders using recent results in generalised geometry~\cite{AGGPW16}. We will give explicit examples for S$^{7}$, Q$^{1,1,1}$ and M$^{1,1,1}$. A similar approach was taken recently to analyse marginal deformations of CFTs dual to $\text{AdS}^{5}\times M$ solutions in type IIB, where $M$ is Sasaki–Einstein~\cite{AGGPW16} – this paper applies the same idea to AdS$_{4}$ solutions in M-theory.

Our analysis applies to any Sasaki–Einstein solution preserving at least eight supercharges. We find that the marginal deformations are encoded in a function of weight four under the Reeb vector that is holomorphic on the Calabi–Yau cone over the Sasaki–Einstein. For example, for $\text{S}^{7}$ we find the marginal deformations are defined by a quartic function of the usual complex coordinates $z_{i}$ on $\mathbb{C}^{4}$. Such a quartic function generically has 35 complex degrees of freedom. The obstruction appears in our formalism as an extra symplectic quotient that reduces this to 20 complex degrees of freedom, agreeing with the counting from the dual field theory.

The reason for using generalised geometry is its packaging of supergravity degrees of freedom in the same way as the dual field theory. The supergravity solution is characterised by a pair of geometric structures – a hypermultiplet (H) structure and a vector-multiplet (V) structure – that together define an exceptional Sasaki–Einstein (ESE) structure~\cite{AW15b}. This is analogous to the complex and symplectic structures that describe a Calabi–Yau geometry. The H structure corresponds to hypermultiplet degrees of freedom in the gauged supergravity one would find by compactifying on the background. Deformations of the H structure map to superpotential deformations in the dual CFT. If the perturbed AdS$_{4}$ background is supersymmetric, the superpotential deformation is exactly marginal. If we restrict to first-order deformations, it is equivalent to finding the marginal deformations of the CFT. The new AdS$_{4}$ solution is supersymmetric provided the geometric structures are integrable. This integrability appears in our formalism as the vanishing of a triplet of moment maps. Given a first-order solution to the moment maps, there can be obstructions to extending to higher orders. We will see that this obstruction is precisely the condition for a marginal deformation to be exactly marginal.

We will also comment on how the all-orders solutions of Lunin and Maldacena appear in our set up. Focussing on $\text{AdS}_{4}\times\text{S}^{7}$, we will see that the new solution can be encoded in generalised geometry by the action of a tri-vector. The Lunin–Maldacena (LM) solution has been previously understood for $\AdS5$ backgrounds in type IIB by considering the action of a bi-vector on the pure spinors that describe the generalised complex structure of the solution~\cite{MPZ06}. One cannot repeat this analysis for $\AdS4$ solutions in M-theory as the backgrounds are not characterised by a pair of pure spinors. More recent work~\cite{BKO+18,BCS+18}, inspired by the open-closed string map~\cite{SW99}, has suggested that any supergravity solution with isometries and a vanishing $B$ field admits a bi-vector deformation (though not necessarily preserving supersymmetry). The bi-vector is formed from antisymmetric products of Killing vectors with constant coefficients, where the coefficients give an $r$-matrix solution to the classical Yang–Baxter equation. The deformed metric and $B$ field are then extracted from a matrix inversion. This approach will not work for M-theory without extensions that do not rely on underlying stringy properties. Instead, the LM solutions must be understood using the full formalism of exceptional generalised geometry. We hope to return to the question of understanding the other deformations in this formalism in the near future.

We begin in section \ref{sec:Sasaki-Einstein-in-M-theory} with a discussion of seven-dimensional Sasaki–Einstein manifolds as backgrounds in M-theory. We then give details on $\Ex{7(7)}\times\mathbb{R}^{+}$ generalised geometry and review how conventional Sasaki–Einstein solutions can be reformulated in terms of exceptional Sasaki–Einstein structures. In section \ref{sec:Linearised-deformations}, we find the linearised deformations of this structure and give the four-form flux generated by the deformation. The expression for the flux is valid for any Sasaki–Einstein background and includes the linearised fluxes found using the solution-generating of LM technique as a special case. We then explain how the exactly marginal deformations are picked out by an obstruction condition. In section \ref{sec:Examples} we look at the examples of S$^{7}$ and Q$^{1,1,1}$, and find agreement with the known results. Finally, in section \ref{sec:The-Lunin=002013Maldacena-background} we comment on how the all-orders solutions of LM can be described by a tri-vector deformation.

\section{Sasaki–Einstein backgrounds in generalised geometry\label{sec:Sasaki-Einstein-in-M-theory}}

We begin with a review of $\AdS4\times M$ solutions in eleven-dimensional supergravity where the internal space $M$ is Sasaki–Einstein. We then give an outline of exceptional Sasaki–Einstein (ESE) structures in $\Ex{7(7)}\times\mathbb{R}^{+}$ generalised geometry that describe general flux backgrounds with a seven-dimensional internal space, show how the Sasaki–Einstein solutions embed in these structures. It is these structures that we deform in section \ref{sec:Linearised-deformations}.

\subsection{Sasaki–Einstein solutions in M-theory}

Backgrounds of the form $\text{AdS}_{4}\times M$, where $M$ is Sasaki–Einstein, are supersymmetric solutions of eleven-dimensional supergravity preserving at least eight supercharges~\cite{AFHS99}. They are dual to the three-dimensional superconformal field theory living on a stack of M2-branes placed at the tip of the metric cone over $M$. The eleven-dimensional metric is a product
\begin{equation}
\dd s^{2}=\tfrac{1}{4}\dd s^{2}(\textnormal{\text{AdS}}_{4})+\dd s^{2}(M),
\end{equation}
where $M$ is seven dimensional and we set the AdS radius to $1$.\footnote{The AdS radius is $1$ with respect to $\dd s^{2}(\textnormal{\text{AdS}}_{4})$. The explicit factor of $1/4$ takes care of the usual normalisation.} The solution is supersymmetric for four-form flux proportional to the volume form on $\text{AdS}_{4}$~\cite{GKVW09}
\begin{equation}
\dd A=F=\tfrac{3}{8}\vol(\text{AdS}_{4}).
\end{equation}
Restricting the fields to $M$ gives a seven-form flux on the Sasaki–Einstein that is dual to the four-form flux in eleven-dimensions, given by\footnote{Our conventions for the Hodge star are given in appendix A of \cite{AW15}.}
\begin{equation}
\dd\tilde{A}=\tilde{F}=-6\vol_{7}.
\end{equation}

Sasaki–Einstein spaces posses a nowhere-vanishing vector field $\xi$, known as the Reeb vector. If the Sasaki–Einstein space is regular, $\xi$ defines a $\Uni{1}$ fibration over a K\"ahler–Einstein base $M_{6}$, so the metric can be written as
\begin{equation}
\dd s^{2}(M)=\sigma^{2}+\dd s^{2}(M_{6}),
\end{equation}
where $\sigma$, the one-form dual to $\xi$, is known as the contact form. Sasaki–Einstein manifolds admit an $\SU{3}$ structure defined by a complex three-form $\Omega$, a real two-form $\omega$ and the contact form $\sigma$. The $\SU3$ structure implies $\{\Omega,\omega,\sigma\}$ satisfy a set of algebraic conditions
\begin{equation}
\frac{\ii}{8}\Omega\wedge\bar{\Omega}=\frac{1}{3!}\omega\wedge\omega\wedge\omega,\qquad\imath_{\xi}\Omega=\imath_{\xi}\omega=0,\qquad\imath_{\xi}\sigma=1.\label{eq:SE_algebraic}
\end{equation}
The intrinsic torsion of the $\SU3$ structure is characterised by a number of torsion classes~\cite{BCL06,DP04}, or equivalently differential conditions on $\{\Omega,\omega,\sigma\}$:
\begin{equation}
\dd\omega=0,\qquad\dd\Omega=4\ii\,\sigma\wedge\Omega,\qquad\dd\sigma=2\,\omega.\label{eq:SE_diff}
\end{equation}
The Reeb vector $\xi$ is a Killing vector that preserves $\sigma$ and $\omega$ but rotates $\Omega$ by a phase
\begin{equation}
\mathcal{L}_{\xi}\sigma=\mathcal{L}_{\xi}\omega=\mathcal{L}_{\xi}g=0,\qquad\mathcal{L}_{\xi}\Omega=4\ii\,\Omega.\label{eq:SE_Lie}
\end{equation}
The rotation of $\Omega$ corresponds to the R-symmetry of the $\mathcal{N}=2$ solution. Note that we can always find an orthonormal frame $\{e^{a}\}$ on $M$ in which these objects are given by
\begin{equation}
\Omega=(e^{1}+\ii\,e^{2})\wedge(e^{3}+\ii\,e^{4})\wedge(e^{5}+\ii\,e^{6}),\qquad\omega=e^{12}+e^{34},\qquad\sigma=e^{7},\label{eq:SE_frame}
\end{equation}
where we are using the shorthand $e^{1}\wedge e^{2}\equiv e^{12}$. This choice of $\Omega$ satisfies $\bar{\Omega}^{\sharp}\lrcorner\Omega=8$ where $\bar{\Omega}^{\sharp}$ is given by raising the indices of $\bar{\Omega}$ with the metric. The invariant tensors define a volume as 
\begin{equation}
\vol_{7}=e^{1234567}=\frac{1}{3!}\sigma\wedge\omega\wedge\omega\wedge\omega.
\end{equation}
Raising an index on $\omega$ defines a complex structure $I$ on the space transverse to $\xi$. Explicitly it is given by
\begin{equation}
I=-\omega=\frac{\ii}{4}(j\bar{\Omega}{}^{\sharp}\lrcorner j\Omega-j\Omega^{\sharp}\lrcorner j\bar{\Omega}),\qquad I\cdot\Omega=-3\ii\,\Omega,
\end{equation}
where the $j$ operator and $\cdot$, the $\gl7$ adjoint action on a tensor, are defined in \cite{CSW11}.

We are interested in AdS$_{4}$ solutions dual to marginal deformed 3d $\mathcal{N}=2$ CFTs. These deformed solutions no longer admit Sasaki–Einstein metrics so we cannot simply look for solutions to $\dd\delta\omega=0$, and so on. There are two paths one could take. The first path involves using the supersymmetry variations and deforming the Killing spinors, similar to what was done for $\AdS5\times\text{S}^{5}$ in type IIB in \cite{AKY02} up to third order in the deformation. Unfortunately, this method does not give much insight into what is happening geometrically and it requires explicit coordinates on the internal space – one would have to repeat this analysis for each Sasaki–Einstein. The second path tries to use as much geometric structure as possible. The idea is to recast a general $\AdS4$ flux background in M-theory in terms of invariant objects and then embed the Sasaki–Einstein structure inside these. The deformed solutions can still be represented as one of these more general structures and in principle one can do this for all Sasaki–Einstein solutions at once. This is the path we will take.

\subsection{Generalised geometry}

The generalisations of $\Omega$, $\omega$ and $\sigma$ that describe general $\AdS4$ flux backgrounds in M-theory are invariant objects in $\Ex{7(7)}\times\mathbb{R}^{+}$ generalised geometry. Generalised geometry is the study of structures on a generalised tangent bundle $E$~\cite{Hitchin02,Gualtieri04,Hull07,PW08}. For M-theory on a seven-dimensional manifold $M$, the generalised tangent bundle is locally given by
\begin{equation}
E\simeq TM\oplus\ext^{2}T^{*}M\oplus\ext^{5}T^{*}M\oplus(T^{*}M\otimes\ext^{7}T^{*}M).
\end{equation}
This is an $\Ex{7(7)}\times\mathbb{R}^{+}$ vector bundle with fibres transforming in the $\rep{56}_{1}$ representation.\footnote{An $\Ex{7(7)}\times\mathbb{R}^{+}$ scalar of weight $k$ is denoted by $\rep{1}_{k}$ and is a section of $(\det T^{*}M)^{k/2}$. Under a $\GL{7;\mathbb{R}}$ transformation $r$ such an object transforms with $(\det r)^{-k/2}$.} Globally, $E$ is defined by a series of extensions which allow it to be patched together using diffeomorphisms and gauge transformations of a three- and six-form potential. The generalised frame bundle $\tilde{F}$ is an $\Ex{7(7)}\times\mathbb{R}^{+}$ principal bundle constructed from frames for $E$~\cite{CSW11,CSW14}. We also need the adjoint bundle $\ad\tilde{F}$, which has a decomposition in terms of $\GL{7;\mathbb{R}}$ tensor bundles as
\begin{equation}
\ad\tilde{F}\simeq\mathbb{R}\oplus(TM\oplus T^{*}M)\oplus\ext^{3}T^{*}M\oplus\ext^{6}T^{*}M\oplus\ext^{3}TM\oplus\ext^{6}TM.
\end{equation}
This is an $\Ex{7(7)}\times\mathbb{R}^{+}$ bundle with fibres transforming in the $\rep{133}_{0}$ representation.

The generalised tangent bundle admits a generalisation of the Lie derivative that encodes the bosonic symmetries of supergravity. Given a generalised vector field $V\in\Gamma(E)$, one can define the action of the Dorfman derivative (or generalised Lie derivative) $\Dorf_{V}$ on a generalised tensor. This endows $E$ with the structure of a Leibniz algebroid. 

In what follows we will need to write sections of these bundle and use explicit maps such as the adjoint action $\ad\tilde{F}\times E\rightarrow E$. We will also need expressions for the Dorfman derivative and a number of invariant forms. For these we follow the conventions laid out in \cite{CSW11,AW15}.

\subsubsection*{Generalised $G$-structures}

In conventional geometry one can define a $G$-structure as a reduction of the structure group of the frame bundle of a $d$-dimensional manifold from $\GL{d;\mathbb{R}}$ to $G$. This can equivalently be described by a set of a nowhere-vanishing $G$-invariant tensors. For example, we saw in the previous section that the $\SU3$ structure of a seven-dimensional Sasaki–Einstein manifold is characterised by $\{\Omega,\omega,\sigma\}$. One can check that the expressions for these objects given in (\ref{eq:SE_frame}) is indeed invariant under $\SU3\subset\GL{7;\mathbb{R}}$ rotations of the frame $\{e^{a}\}$. 

One can do the same in generalised geometry. A generalised $G$-structure is a reduction of the structure group of $\tilde{F}$ from $\Ex{7(7)}\times\mathbb{R}^{+}$ to some subgroup $G$. Again it can be characterised by a set of $G$-invariant tensors. Let us introduce two such generalised $G$-structures.

A \emph{hypermultiplet} or \emph{H} structure defines a reduction of the structure group of $\tilde{F}$ to $\text{Spin}^{*}(12)$~\cite{AW15,GLSW09}. The invariant tensors are an $\SU{2}$ triplet of objects $J_{\alpha}$ transforming in the $\rep{133}_{1}$ of $\Ex{7(7)}\times\mathbb{R}^{+}$ with a non-zero weight under an overall $\mathbb{R}^{+}$ scaling. The $\SU{2}$ corresponds to the R-symmetry of the $\mathcal{N}=2$ solution. The $J_{\alpha}$ satisfy an $\SU{2}$ algebra
\begin{equation}
[J_{\alpha},J_{\beta}]=2\kappa\,\epsilon_{\alpha\beta\gamma}J_{\gamma},\label{eq:su2_algebra}
\end{equation}
where $\kappa$ is a section of $(\det T^{*}M)^{1/2}$ and $\kappa^{2}$ is the $\Ex{7(7)}$-invariant volume, given by $\kappa^{2}=\vol_{7}$ for backgrounds with a vanishing warp factor. The norms of the $J_{\alpha}$ calculated using the $\ex{7(7)}$ Killing form, which we denote by $\tr$, are fixed to
\begin{equation}
\tr(J_{\alpha},J_{\beta})=-\kappa^{2}\delta_{\alpha\beta}.\label{eq:J_a_norm}
\end{equation}

A \emph{vector-multiplet} or \emph{V} structure defines a reduction of the structure group of $\tilde{F}$ to $\Ex{6(2)}$. As the scalars of vector multiplets are complex, we parametrise the structure by $X=K+\ii\,\hat{K}$ transforming in $\rep{56}_{1}^{\mathbb{C}}$. $X$ is invariant under an $\Ex{6(2)}$ subgroup provided
\begin{equation}
s(K,\hat{K})=2\,\sqrt{q(K)}>0,\label{eq:e7_comp}
\end{equation}
where $s$ and $q$ are the $\Ex{7(7)}$ symplectic and quartic invariants. Note that $K$ and $\hat{K}$ are not independent – one can construct $\hat{K}$ from $K$ using a Hitchin functional, similar to recovering $\Omega$ from its real part as described in \cite{Hitchin00}, so $K$ on its own is enough to define the V structure.\footnote{The object $\hat{K}$ has appeared in work on black hole entropy and U-duality, where it is known as the Freudenthal dual of $K$~\cite{BDD+09,LS12,BDF+13}.}

If the H and V structures are \emph{compatible}, they intersect on an $\SU{6}$ structure 
\begin{equation}
\text{Spin}^{*}(12)\cap\Ex{6(2)}=\SU{6}.
\end{equation}
The H and V structures must satisfy compatibility conditions to ensure they are invariant under a common $\SU6$ subgroup
\begin{equation}
J_{\alpha}\cdot K=0,\qquad s(K,\hat{K})=\kappa^{2},\label{eq:e7_comp_condition}
\end{equation}
where $\cdot$ is the adjoint action $\rep{133}\times\rep{56}\rightarrow\rep{56}$. Notice that these are analogous to $\omega\wedge\Omega=0$ and $|\Omega|^{2}\propto\omega^{3}$ for a conventional $\SU{3}$ structure in six dimensions. 

The existence of an $\SU{6}$ structure implies that compactification on such a background would give an $\mathcal{N}=2$ effective theory. However we are interested in true supersymmetric solutions to the eleven-dimensional equations of motion. For this we require the $\SU{6}$ structure to be \emph{integrable} or \emph{torsion-free}. This amounts to a set of differential conditions on the invariant objects $\{J_{\alpha},K,\hat{K}\}$ which give the generalisation of (\ref{eq:SE_diff}) for flux backgrounds (see \cite{AW15b} for more details). Integrability of the H structure is given by 
\begin{equation}
\mu_{\alpha}(V)\coloneqq-\tfrac{1}{2}\epsilon_{\alpha\beta\gamma}\int_{M}\tr(J_{\beta},\Dorf_{V}J_{\gamma})=\lambda_{\alpha}\int_{M}s(V,\hat{K})=2\,\lambda_{\alpha}\int_{M}\sqrt{q(V,K,K,K)},\label{eq:moment_maps}
\end{equation}
where the $\lambda_{3}=4$ and $\lambda_{1,2}=0$ are related to the cosmological constant. Integrability of the V structure is given by
\begin{equation}
\Dorf_{K}K=0.\label{eq:K_cond}
\end{equation}
The integrability conditions for an $\SU{6}$ structure are the above conditions plus\footnote{Note that we have used the freedom to choose the phase of $X=K+\ii\,\hat{K}$ in writing these equations.}
\begin{equation}
\Dorf_{K}J_{\alpha}=\epsilon_{\alpha\beta\gamma}\lambda_{\beta}J_{\gamma},\qqq\Dorf_{\hat{K}}J_{\alpha}=0.\label{eq:u1_charge}
\end{equation}

The appearance of $\SU{6}$ fits with $\mathcal{N}=2$ supersymmetry – one can see this by thinking about the Killing spinors. One can always find an torsion-free $\SU{8}$ structure, corresponding to the existence of a generalised metric~\cite{CSW11,CSW14}. The supersymmetry parameters in the Killing spinor equations then transform as the $\rep{8}$ of this $\SU{8}$. A choice of two orthogonal spinors is invariant under a reduced $\SU{6}\subset\SU{8}$ subgroup, so that a torsion-free $\SU{6}$ structure gives rise to $\mathcal{N}=2$ supersymmetry. 

We have reviewed ESE structures in generalised geometry and given the integrability conditions that are equivalent to the preservation of eight supercharges. We now show how AdS$_{4}$ backgrounds with a Sasaki–Einstein factor in M-theory can be embedded as an ESE structure.

\subsection{Embedding as an ESE structure}

For Sasaki–Einstein solutions, $K$ generalises the contact structure and $J_{\alpha}$ generalises the complex structure. Thus, we expect $\xi$, $\sigma$ and $\omega$ to appear in $K$, and $\Omega$ and $I$ to appear in $J_{\alpha}$. This embedding was first given in \cite{AW15b}.

The H structure $J_{\alpha}$ is given by
\begin{equation}
\begin{split}J_{+} & =\tfrac{\kappa}{2}\Omega-\tfrac{\kappa}{2}\Omega^{\sharp},\\
J_{3} & =\tfrac{\kappa}{2}I-\tfrac{\kappa}{2}\tfrac{\ii}{8}\Omega\wedge\bar{\Omega}-\tfrac{\kappa}{2}\tfrac{\ii}{8}\Omega^{\sharp}\wedge\bar{\Omega}^{\sharp},
\end{split}
\label{eq:H_structure}
\end{equation}
where $J_{\pm}=J_{1}\pm\ii\,J_{2}$. One can check using the expressions in \cite{AW15} that these satisfy the $\SU{2}$ algebra (\ref{eq:su2_algebra}) and the normalisation conditions (\ref{eq:J_a_norm}), where the $\Ex{7(7)}$-invariant volume is 
\begin{equation}
\kappa^{2}=\vol_{7}.
\end{equation}
The V structure is given by
\begin{equation}
X=K+\ii\,\hat{K}=\xi+\ii\,\omega-\tfrac{1}{2}\sigma\wedge\omega\wedge\omega-\ii\,\sigma\otimes\vol_{7}.\label{eq:V_structure}
\end{equation}
Again one can check that this is normalised according to (\ref{eq:e7_comp}). We can understand why the vector-multiplet structure embeds in this fashion. Locally a Sasaki–Einstein is $M_{6}\times\text{S}^{1}$, where $M_{6}$ is K\"ahler–Einstein and so has an $\SU{3}$ structure. By removing the dependence of the fibre direction from $X$, we recover the usual pure spinor corresponding to the symplectic form, $\ee^{\ii\,\omega}$. Finally, one can show that $J_{\alpha}$ and $K$ satisfy the compatibility conditions (\ref{eq:e7_comp_condition}) so that the $\text{Spin}^{*}(12)$ and $\Ex{6(2)}$ structures intersect on a common $\SU{6}$ structure.

Note that we have not included the action of $\ee^{\tilde{A}}$, the potential for the dual seven-form flux $\tilde{F}$, in $J_{\alpha}$ or $X$. As discussed in \cite{AW15,AW15b,AGGPW16}, we are free to include this by using a twisted Dorfman derivative $\hat{\Dorf}$ in the differential conditions that follow. In Hitchin's generalised geometry, this is analogous to working with $\dd_{H}=\dd-H\wedge$ and the undressed pure spinors $\Phi_{\pm}$ instead of $\dd$ and $\ee^{-B}\Phi_{\pm}$.

We have reviewed seven-dimensional Sasaki–Einstein backgrounds in M-theory and how they can be rephrased as ESE structures in $\Ex{7(7)}\times\mathbb{R}^{+}$ generalised geometry. We now want to investigate the possible deformations of this structure that are still integrable. In other words, we look for deformations of the supergravity solution that preserve eight supercharges. We expect these to be dual to exactly marginal deformations in the field theory.

\section{Linearised deformations\label{sec:Linearised-deformations}}

The Sasaki–Einstein background is defined by a pair of structures $K$ and $J_{\alpha}$ which define a torsion-free $\SU{6}$ structure. $K$ and $J_{\alpha}$ each transform under $\Ex{7(7)}\times\mathbb{R}^{+}$ and so a choice of $\{J_{\alpha},K\}$ is a point on an orbit of the $\Ex{7(7)}\times\mathbb{R}^{+}$ action. A linearised deformation of these structures is given by moving to a point on the orbit that corresponds to the deformed background, so linearised deformations can be parametrised by elements $\mathcal{A}$ in the adjoint of $\Ex{7(7)}\times\mathbb{R}^{+}$. The marginal deformations of the dual $\mathcal{N}=2$ SCFT then correspond to deformations of $\{J_{\alpha},K\}$ that satisfy the supersymmetry conditions. We will solve the supersymmetry conditions in two steps: first we solve the linearised moment map conditions, then we impose conditions (\ref{eq:u1_charge}) which require the adjoint element which parametrises the deformation to be uncharged under the $\Uni{1}_{\text{R}}$ generated by $K$.

As discussed in \cite{GKSTW10}, in a three-dimensional $\mathcal{N}=2$ SCFT the marginal superpotential deformations are given by turning on chiral primary operators of superfield dimension two. Marginal deformations of K\"ahler type are given by real primary operators of dimension one. These operators are conserved currents and so do not deform the Lagrangian, implying there are no marginal deformations of K\"ahler type. The bulk picture for these deformations is given by deforming $J_{\alpha}$ and $K$. The deformations of $\{J_{\alpha},K\}$ come in two types:\footnote{One can check that the third type of deformation where $J_{\alpha}$ and $K$ are deformed simultaneously does not change the generalised metric; this means that the physical supergravity fields do not change and so it is not a new solution. If this type of deformation is present, it corresponds to a deformation of the Killing spinors and indicates the presence of extra supersymmetries in the undeformed solution.}
\begin{align*}
\text{Kähler deformations:}\quad & \delta K\neq0,\quad\delta J_{\alpha}=0,\\
\text{Superpotential deformations:}\quad & \delta J_{\alpha}\neq0,\quad\delta K=0.
\end{align*}
We identify the deformations of $\{J_{\alpha},K\}$ with operators in the dual SCFT by appealing to the four-dimensional gauged supergravity one would find by compactifying the internal seven dimensions: fluctuations of $J_{\alpha}$ and $K$ are hypermultiplets and vector multiplets in the four-dimensional theory, which correspond to chiral and real primary operators in the CFT.

\subsubsection{No K\"ahler deformations}

Let us start by considering the K\"ahler deformations where $J_{\alpha}$ is fixed and $K$ is deformed. These are deformations which deform the $\SU6$ structure but preserve the $\Spinstar{12}$ structure. First note that the normalisation condition (\ref{eq:e7_comp_condition}) implies that simple rescalings of $K$ will not lead to a supersymmetric solution. In other words, deformations by $\mathbb{R}^{+}$ elements are forbidden, leaving only $\Ex{7(7)}$ elements. Looking back to the moment maps (\ref{eq:moment_maps}), the left-hand side depends only on $J_{\alpha}$ and so does not change. Thus the supersymmetric deformations of K\"ahler type must satisfy
\begin{equation}
\int_{M}\sqrt{q(V,\delta K,K,K)}=0,
\end{equation}
for any generalised vector $V$. Since $q$ is an $\Ex{7(7)}$ invariant we have
\begin{equation}
q(\delta V,K,K,K)+3\,q(V,\delta K,K,K)=0,
\end{equation}
and so the moment map condition becomes
\begin{equation}
\int_{M}\sqrt{q(\delta V,K,K,K)}=0\qquad\forall\,V.
\end{equation}
This vanishes if and only if $\delta V$ has no piece proportional to $K$ for any choice of $V$. Under $\Ex{6(2)}\times\Uni{1}$, the adjoint and vector of $\Ex{7(7)}$ decompose as
\begin{equation}
\begin{split}\rep{133} & \rightarrow\rep{78}_{0}+\rep{27}_{2}+\overline{\rep{27}}_{-2}+\rep{1}_{0},\\
\rep{56} & \rightarrow\rep{27}_{-1}+\overline{\rep{27}}_{1}+\rep{1}_{3}+\rep{1}_{-3},
\end{split}
\end{equation}
where $\rep{78}$ is the adjoint of $\Ex{6(2)}$ and the singlets in the $\rep{56}$ correspond to $X$ and $\bar{X}$. $\delta V$ will have no $K$ component only if the deformation is by an element of the $\rep{78}$ – the other representations will generate the unwanted singlet when acting on an arbitrary generalised vector $V$. However, elements of the $\rep{78}$ stabilise the underlying $\Ex{6(2)}$ structure and so do not give a new V structure. We conclude there are no deformations of K\"ahler type.

\subsection{Superpotential deformations}

The dual of a superpotential deformation is a deformation of the $\SU{6}$ structure that leaves the $\Ex{6(2)}$ structure invariant. The linearised deformations can be parametrised by an element of the adjoint as
\begin{equation}
\delta K=\mathcal{A}\cdot K=0,\qqq\delta J_{\alpha}=[\mathcal{A},J_{\alpha}]\neq0,
\end{equation}
for some $\mathcal{A}\in\Gamma(\ad\tilde{F})$. As they leave the $\Ex{6(2)}$ structure invariant, such deformations parametrise $\Ex{6(2)}/\SU{6}$. The adjoint of $\Ex{6(2)}$ decomposes under $\SU{2}\times\SU{6}$ as
\begin{equation}
\rep{78}\rightarrow(\rep{3},\rep{1})+(\rep{1},\rep{35})+(\rep{2},\rep{20}).
\end{equation}
The first term corresponds to $\SU{2}$ rotations of the triplet $J_{\alpha}$ – such deformations do not change the generalised metric (to first order) and so do not deform the supergravity fields. The second term is the adjoint of $\SU{6}$, which leaves both $J_{\alpha}$ and $K$ invariant by definition. Thus we can take the first-order deformations to be in the $(\rep{2},\rep{20})$, forming a doublet under the $\SU{2}$ defined by $J_{\alpha}$. We are free to organise them to be eigenstates of $J_{3}$
\begin{equation}
[J_{3},\mathcal{A}_{\lambda}]=\ii\lambda\kappa\,\mathcal{A}_{\lambda}.
\end{equation}
The $\lambda=\pm2$ eigenstates correspond to $J_{\mp}$. The deformations we want are the $\lambda=\pm1$ eigenstates. To find their explicit form, it is useful to note that the $\Ex{7(7)}\times\mathbb{R}^{+}$ algebra allows us to split the generic deformation as
\begin{equation}
\mathcal{A}_{+}=a+\alpha,\qqq\tilde{\mathcal{A}}_{+}=g+\tilde{a}+\tilde{\alpha},
\end{equation}
where $\mathcal{\mathcal{A}}_{-}=\mathcal{A}_{+}^{*}$. Notice that an $\tilde{\mathcal{A}}_{-}$-type deformation can be obtained from $\mathcal{A}_{+}$ by acting with the lowering operator $J_{-}$. It turns out that only $\mathcal{A}_{\pm}$ can give marginal deformations so we do not discuss $\tilde{\mathcal{A}}_{\pm}$ further.\footnote{The argument is identical to that in appendix B of \cite{AGGPW16}.}

The $\lambda=1$ eigenstate $\mathcal{A}_{+}$ has three-form and three-vector components given by
\begin{equation}
\begin{split}\mathcal{A}_{+} & =a+\alpha,\\
a & =\phi+f\bar{\Omega}+\sigma\wedge(\nu^{\sharp}\lrcorner\bar{\Omega}),\\
\alpha & =\phi^{\sharp}+f\bar{\Omega}^{\sharp}+\xi\wedge(\bar{\Omega}^{\sharp}\lrcorner\nu),
\end{split}
\label{eq:A+}
\end{equation}
where $\phi$ is of type $(1,2)$ and $\nu$ is $(1,0)$ with respect to the undeformed complex structure $I$. The sharp $\sharp$ indicates we have raised the indices of the form with the undeformed metric so that $\nu^{\sharp}$ is a $(0,1)$-vector, and so on. The element $\mathcal{A}_{+}$ leaves $K$ and $\hat{K}$ invariant provided
\begin{equation}
\begin{split}\bigl(\xi\wedge(\bar{\Omega}^{\sharp}\lrcorner\nu)\bigr)\lrcorner\vol_{7} & =\ii\,\nu\wedge\bar{\Omega},\\
\bar{\Omega}^{\sharp}\lrcorner\vol_{7} & =-\ii\,\sigma\wedge\bar{\Omega},\\
\phi^{\sharp}\lrcorner\vol_{7} & =\ii\,\sigma\wedge\phi.
\end{split}
\label{eq:contractions}
\end{equation}

Deformations that live in $\SU{8}$ leave the generalised metric invariant and so do not change the physical fields of the solution. The condition for an adjoint element with only three-form and three-vector components to be in $\SU8$ is $a^{\sharp}=-\alpha$. Our deformation satisfies $a^{\sharp}=\alpha$, so we see our deformation is not in $\SU{8}$ and will change the physical fields. The following identities will be useful in finding the conditions on $a$ and $\alpha$:
\begin{equation}
\vol_{7}(\beta^{(3)}\lrcorner\beta_{(3)})=(\beta^{(3)}\lrcorner\vol_{7})\wedge\beta_{(3)},\qqq\vol_{7}(\beta^{(6)}\lrcorner\beta_{(6)})=(\beta^{(6)}\lrcorner\vol_{7})\wedge\beta_{(6)},
\end{equation}
where $\beta_{(n)}$ and $\beta^{(m)}$ are arbitrary $n$-forms and $m$-vectors.

\subsubsection{Linearised moment maps}

We now make a first-order deformation of $J_{+}$ by $\mathcal{A}_{+}$ so that
\begin{equation}
\delta J_{+}=[\mathcal{A}_{+},J_{+}],\qqq\delta J_{-}=[\mathcal{A}_{-},J_{-}].
\end{equation}
The corresponding linear deformation of $J_{3}$ is given by
\begin{equation}
\delta J_{3}=[\mathcal{A}_{+}+\mathcal{A}_{-},J_{3}].
\end{equation}
The linearised moment map equations $\delta\mu_{\alpha}(V)=0$ are then
\begin{equation}
\delta\mu_{\alpha}(V)=\int\kappa\,\tr\bigl(J_{\alpha},\hat{\Dorf}_{V}(\mathcal{A}_{+}+\mathcal{A}_{-})\bigr)=0.
\end{equation}
The idea is to expand these conditions using (\ref{eq:A+}) and find the conditions that must be satisfied by $\{f,\phi,\nu\}$.

The variation of $\mu_{3}$ can be simplified
\begin{equation}
\begin{split}\delta\mu_{3}(V) & \propto\int\kappa\tr(J_{3},\hat{\Dorf}_{V}\re\mathcal{A}_{+})\\
 & \propto\int\kappa\tr(J_{3},[\dd\omega,\re\mathcal{A}_{+}])\\
 & \propto\int\kappa\tr(\dd\omega,[J_{3},\re\mathcal{A}_{+}])\\
 & \propto\int\kappa^{2}\tr(\dd\omega,\im\mathcal{A}_{+}),
\end{split}
\end{equation}
where we have used the form of the Dorfman derivative, the properties of the trace and the bracket, and $[J_{3},\mathcal{A}_{\pm}]\propto\pm\mathcal{A}_{\pm}$.\footnote{Here we are thinking of $\dd\omega$ as a section of the adjoint bundle.} Using the form of the trace we find
\begin{equation}
\begin{split}\delta\mu_{3}(V) & \propto\int\kappa^{2}\im\alpha\lrcorner\dd\omega\\
 & \propto\int\dd(\im\alpha\lrcorner\vol_{7})\wedge\omega.
\end{split}
\end{equation}
As this must vanish for any $\omega$, the first-order deformation of $\mu_{3}$ vanishes if
\begin{equation}
\dd(\im\alpha\lrcorner\vol_{7})=0.\label{eq:mu3_cond}
\end{equation}
We have an explicit form for $\mathcal{A}_{+}$ in (\ref{eq:A+}) and the contractions of its components with $\vol_{7}$ given in (\ref{eq:contractions}). Using these and the decomposition into complex type with respect to $I$, (\ref{eq:mu3_cond}) is equivalent to
\begin{equation}
\begin{split}\partial\phi+\bar{\partial}\bar{\phi} & =0,\\
\partial\nu & =0,\\
\bar{\partial}\phi-\partial f\wedge\bar{\Omega}-\mathcal{L}_{\xi}\nu\wedge\bar{\Omega}+4\ii\,\nu\wedge\bar{\Omega} & =0.
\end{split}
\label{eq:mu3}
\end{equation}
It is simple to repeat this procedure for $\delta\mu_{+}=0$, from which we find
\begin{equation}
\begin{split}\partial\nu & =0,\\
\bar{\partial}\nu & =2\,f\omega,\\
\mathcal{L}_{\xi}\nu & =-\partial f,\\
\bar{\partial}f & =0,\\
\bar{\partial}(\nu\lrcorner\bar{\Omega})+6\ii\,f\bar{\Omega} & =0,\\
-2\ii\,\bar{\partial}\phi-2\ii\,\omega\wedge(\nu\lrcorner\bar{\Omega})+6\,\nu\wedge\bar{\Omega}+\ii\,\partial f\wedge\bar{\Omega}+\ii\,\mathcal{L}_{\xi}\nu\wedge\bar{\Omega} & =0.
\end{split}
\label{eq:mu+}
\end{equation}
These are the conditions that $f$, $\phi$ and $\nu$ must satisfy for $J_{\alpha}+\delta J_{\alpha}$ to define a supersymmetric solution to first order in the deformation parameter. Note that we can simplify some of the relations using
\begin{equation}
\nu\wedge\bar{\Omega}+\ii\,\omega\wedge(\nu^{\sharp}\lrcorner\bar{\Omega})=0,
\end{equation}
which is valid for any $(1,0)$-form $\nu$.

\subsubsection{Marginal deformations}

The moment maps are only a subset of the full supersymmetry conditions we need to satisfy to guarantee an $\mathcal{N}=2$ background; we also need to solve the linearised versions of (\ref{eq:K_cond}) and (\ref{eq:u1_charge}). Recall that the superpotential deformations do not change $K$ at first order, so (\ref{eq:K_cond}) is automatically solved. The linearised form of (\ref{eq:u1_charge}) implies $\mathcal{A}_{+}$ should be charge zero under the $\Uni{1}_{\text{R}}$ generated by $K$ (or equivalently the Reeb vector, $\xi$)
\begin{equation}
\hat{\Dorf}_{K}\mathcal{A}_{+}\equiv\mathcal{L}_{\xi}\mathcal{A}_{+}=0.\label{eq:R-charge}
\end{equation}
Without imposing this condition we have found a tower of supergravity modes that are distinguished by their $\Uni{1}_{\text{R}}$ charge. In the dual field theory, the chiral operators have conformal dimensions proportional to their R-charge, so our tower of modes gives a tower of operators that preserve $\mathcal{N}=2$ supersymmetry but break the conformal symmetry.

Combining the differential conditions (\ref{eq:mu3}) and (\ref{eq:mu+}) from the linearised moment maps with the vanishing of the R-charge (\ref{eq:R-charge}) gives the following solution: the components of $\mathcal{A}_{+}$ are determined by the function $f$ that appears in (\ref{eq:A+})
\begin{equation}
\bar{\partial}f=0,\qquad\mathcal{L}_{\xi}f=4\ii f,\qquad\nu=\frac{\ii}{4}\partial f,\qquad\phi=\frac{1}{24}\partial(\partial f\lrcorner\bar{\Omega}).
\end{equation}
The first two conditions means the function $f$ is holomorphic on the cone over $M$ and has charge $+4$ under the Reeb vector, so that $\mathcal{A}_{+}$ is uncharged. One can check this solves the conditions in (\ref{eq:mu3}) and (\ref{eq:mu+}) using
\begin{align}
\bar{\partial}(\partial f\lrcorner\bar{\Omega}) & =-24\,f\bar{\Omega},\\
\partial^{2}=\bar{\partial}^{2} & =0,\\
\partial\bar{\partial}+\bar{\partial}\partial & =-2\,\omega\wedge\mathcal{L}_{\xi},
\end{align}
where the first identity is given in \cite{EST13,ES15}. Note that one can also include an uncharged $\dd$-closed $(1,2)$-form in $\phi$. If this form is exact it is simply a gauge transformation, so whether this shift is an extra degree of freedom depends on the cohomology of $M$.

\subsubsection*{Four-form flux $F=\dd A$}

By considering the effect of a first-order $\mathcal{A}_{+}$ deformation on the generalised metric, one can show that it will turn on three-form potential of the form
\begin{equation}
A=\re\mathcal{A}_{+}.
\end{equation}
If we restrict to marginal deformations that are uncharged under the Reeb vector, we can rewrite this as
\begin{equation}
A=\re\bigl(2\,f\bar{\Omega}+\tfrac{1}{4}\ii\,\sigma\wedge(\partial f^{\sharp}\lrcorner\bar{\Omega})+\dd(\partial f^{\sharp}\lrcorner\bar{\Omega})\bigr),
\end{equation}
where $\partial f^{\sharp}$ is the vector obtained by raising the indices of the one-form $\partial f$ with the undeformed metric on $M$. The four-form flux $F$ due to $A$ is
\begin{equation}
F=\dd\re(2\,f\bar{\Omega}+\tfrac{1}{4}\ii\,\sigma\wedge(\partial f\lrcorner\bar{\Omega})).\label{eq:flux}
\end{equation}
This expression for the flux is valid for the marginal deformations of \emph{any }Sasaki–Einstein background and, as we will show in section \ref{sec:Examples}, it matches the linearised fluxes from the solution-generating technique of Lunin and Maldacena~\cite{LM05}.

\subsection{Obstruction}

Our linearised analysis has given us the supergravity modes that are dual to marginal operators in the SCFT. Ideally we would like to identify which deformations can be extended to all orders in the perturbation. In the field theory, this is equivalent to identifying the subset of marginal operators that are exactly marginal. As we discussed in the introduction, in the CFT this is intimately related to the existence of global symmetries~\cite{GKSTW10}. The supergravity picture of this for a general flux solution was described in \cite{AGGPW16} – let us review this.

Not all of the deformations defined by the holomorphic function $f$ will extend to higher orders as we have to be able to correct the deformation at each order. Let the deformation parameter be $\gamma$. A second-order calculation will show that the differential conditions are not satisfied, but we can add a term of order $\gamma^{2}$ to our original deformation to correct for this, ensuring the deformation still gives a supersymmetric background. The same problem will occur at third order and so on. The question is then which deformations (parametrised by $f$) can be corrected to all orders?

Naively, you have to do the explicit calculation and check whether the solution can be corrected, however the CFT result implies that this is not necessary. On the supergravity side, the trick is to use the formulation of the supersymmetry conditions in terms of moment maps. If one starts at a \emph{generic} point in the space of solutions, there are no obstructions to extending a first-order solution to an all-orders solution. In other words, all marginal deformations are exactly marginal. The case where this argument fails is when the original solution you are deforming is not generic. If the undeformed solution has extra symmetries, the moment maps miss the corresponding constraints and one has an obstruction to extending the deformation past first order. These extra symmetries are a subgroup of $\text{GDiff}_{K}$ – diffeomorphisms and gauge transformations that preserve $K$. The algebra $\mathfrak{gdiff}_{K}$ is parametrised by generalised vectors $V$ that satisfy
\begin{equation}
\mathfrak{gdiff}_{K}=\{V\,\,|\,\,\hat{\Dorf}_{V}K=0\}.
\end{equation}
The subgroup of $\text{GDiff}_{K}$ that also preserves $J_{\alpha}$ is
\begin{equation}
\mathfrak{g}=\{V\in\mathfrak{gdiff}_{K}\,\,|\,\,\hat{\Dorf}_{V}J_{\alpha}=0\}.
\end{equation}
The moment maps vanish identically for $V\in\mathfrak{g}$ (as $\hat{\Dorf}_{V}J_{\alpha}=0$) and so you miss the constraints you would have found from solving $\delta\mu_{\alpha}(V)=0$. What do these $V$ describe? Since the generalised metric $G$ can be obtained from $J_{\alpha}$ and $K$ (see section \ref{sec:Generalised-metric}), the vectors $V$ that generate $\mathfrak{g}$ also preserve the generalised metric
\begin{equation}
\hat{\Dorf}_{V}G=0\quad\forall\,V\in\mathfrak{g}.
\end{equation}
As shown in \cite{AW15b,CS16}, this implies that the vector components of $V$ are conventional Killing vectors for the undeformed metric $g$ and so define its isometry group. Playing this story backwards, we see a global symmetry of the undeformed solution gives generalised vectors for which the moment maps vanish trivially, leading to an obstruction to extending a first-order deformation to all orders. The all-orders deformations are given by finding the first-order deformations – in our case parametrised by a charge +4 holomorphic function $f$ – and then taking a symplectic quotient by the action of the global symmetry group broken by the deformation. This matches the field theory analysis of \cite{GKSTW10}, where the moment map for this quotient will reproduce the one-loop beta function of the dual CFT.

\section{Examples\label{sec:Examples}}

In the previous section we found the first-order fluxes that are dual to marginal deformations for any Sasaki–Einstein background. We now give the explicit examples of supergravity backgrounds where the internal space is $\text{S}^{7}$, $\text{Q}^{1,1,1}$ or $\text{M}^{1,1,1}$. In what follows, it proves useful to take an orthonormal frame on $M$ in which the invariant objects defining the Sasaki–Einstein structure are given by (\ref{eq:SE_frame}).

\subsection{$\text{S}^{7}$}

As an AdS$_{4}$ background in M-theory, the seven-sphere S$^{7}$ preserves 32 supercharges. When viewed as a Sasaki–Einstein manifold, we pick out eight of these supercharges – it is these supercharges that will be preserved by the first-order flux we have given. We can view S$^{7}$ as a $\Uni1$ fibration over $\mathbb{CP}^{3}$, a six-dimensional K\"ahler–Einstein base. The metric on S$^{7}$ can be written as\footnote{We are using $s_{\alpha}$ and $c_{\alpha}$ as shorthand for $\sin\alpha$ and $\cos\alpha$, and similarly for $\beta$ and $\theta$.}~\cite{LM05}
\begin{equation}
\dd s^{2}(\text{S}^{7})=\dd\theta^{2}+s_{\theta}^{2}\,\dd\alpha^{2}+s_{\theta}^{2}s_{\alpha}^{2}\,\dd\beta^{2}+c_{\theta}^{2}\,\dd\phi_{1}^{2}+s_{\theta}^{2}c_{\alpha}^{2}\,\dd\phi_{2}^{2}+s_{\theta}^{2}s_{\alpha}^{2}c_{\beta}^{2}\,\dd\phi_{3}^{2}+s_{\theta}^{2}s_{\alpha}^{2}s_{\beta}^{2}\,\dd\phi_{4}^{2}.\label{eq:S7_metric}
\end{equation}
We introduce an explicit frame in terms of the coordinates on $\text{S}^{7}$:
\begin{equation}
\begin{split}e^{1}+\ii\,e^{2} & =\ee^{4\ii\psi/3}(\dd\theta-\ii s_{\theta}c_{\theta}\,\dd\phi_{1}+\ii s_{\theta}c_{\theta}c_{\alpha}^{2}\,\dd\phi_{2}+\ii s_{\theta}c_{\theta}c_{\beta}^{2}s_{\alpha}^{2}\,\dd\phi_{3}+\ii s_{\theta}c_{\theta}s_{\alpha}^{2}s_{\beta}^{2}\,\dd\phi_{4}),\\
e^{3}+\ii\,e^{4} & =\ee^{4\ii\psi/3}(s_{\theta}\,\dd\alpha-\ii s_{\alpha}c_{\alpha}s_{\theta}\,\dd\phi_{2}+\ii s_{\alpha}c_{\alpha}s_{\theta}c_{\beta}^{2}\,\dd\phi_{3}+\ii s_{\alpha}c_{\alpha}s_{\theta}s_{\beta}^{2}\,\dd\phi_{4}),\\
e^{5}+\ii\,e^{6} & =\ee^{4\ii\psi/3}(s_{\alpha}s_{\theta}\,\dd\beta-\ii s_{\beta}c_{\beta}s_{\theta}s_{\alpha}\,\dd\phi_{3}+\ii s_{\alpha}s_{\beta}c_{\beta}s_{\theta}\,\dd\phi_{4}),\\
e^{7} & =c_{\theta}^{2}\,\dd\phi_{1}+s_{\theta}^{2}c_{\alpha}^{2}\,\dd\phi_{3}+s_{\alpha}^{2}s_{\theta}^{2}c_{\beta}^{2}\,\dd\phi_{3}+s_{\theta}^{2}s_{\alpha}^{2}s_{\beta}^{2}\,\dd\phi_{4},
\end{split}
\label{eq:S7_frame}
\end{equation}
where $4\psi=\phi_{1}+\phi_{2}+\phi_{3}+\phi_{4}$. Using this frame, one can check that the complex, symplectic and contact structures given in (\ref{eq:SE_frame}) satisfy the algebraic and differential conditions (\ref{eq:SE_algebraic}) and (\ref{eq:SE_diff}).

Up to closed three-forms, the marginal deformations are parametrised by a holomorphic function $f$ that descends from the Calabi–Yau cone over $\text{S}^{7}$. The function $f$ is of charge four under the Reeb vector. In our parametrisation, the Reeb vector field is
\begin{equation}
\xi=\partial_{\psi}=\partial_{\phi_{1}}+\partial_{\phi_{2}}+\partial_{\phi_{3}}+\partial_{\phi_{4}}.
\end{equation}
The cone over S$^{7}$ is $\mathbb{C}^{4}$, and the coordinates on S$^{7}$ are related to the usual complex coordinates on $\mathbb{C}^{4}$ by
\begin{equation}
z_{1}=c_{\theta}\ee^{\ii\phi_{1}},\quad z_{2}=s_{\theta}c_{\alpha}\ee^{\ii\phi_{2}},\quad z_{3}=s_{\theta}s_{\alpha}c_{\beta}\ee^{\ii\phi_{3}},\quad z_{4}=s_{\theta}s_{\alpha}s_{\beta}\ee^{\ii\phi_{4}},
\end{equation}
where the coordinates $z_{i}$ have charge +1 under the Reeb vector field
\begin{equation}
\mathcal{L}_{\xi}z_{i}=\ii z_{i}.
\end{equation}
Thus $f$ must be a quartic function of the $z_{i}$. The general form of such a function is
\begin{equation}
f=f^{ijkl}z_{i}z_{j}z_{k}z_{l},
\end{equation}
where $f^{ijkl}$ is a complex symmetric tensor of $\SU4$. There are generically 35 complex degrees of freedom in such a symmetric rank-four tensor, corresponding to the 35 marginal deformations previously discussed by Kol~\cite{Kol02}. 

Requiring our first-order perturbation to extend to higher orders forces us to consider if there are fixed-point isometries at the $\text{S}^{7}$ point in the space of couplings. We can think of $\text{S}^{7}$ as a $\Uni{1}$ fibration over a $\mathbb{CP}^{3}$ base, where the $\SU{4}$ that acts on the base leaves the $\text{S}^{7}$ solution invariant. In other words, $\text{S}^{7}=\SU4/\SU3$ where the action of $\SU4$ preserves the $\Uni1$ fibration – this is not true of the other presentations of $\text{S}^{7}$ as a homogeneous space. This means we have an $\SU{4}$'s worth of fixed-point symmetries, where the marginal deformations defined by $f$ generically break this $\SU4$. To account for this we construct a moment map for the $\SU{4}$ action on the space of couplings and perform a symplectic reduction. The deformations that survive are those that extend to higher orders, namely the exactly marginal deformations. These deformations satisfy
\begin{equation}
f^{iklm}\bar{f}_{jklm}-\tfrac{1}{4}\delta_{\phantom{i}j}^{i}f^{klmn}\bar{f}_{klmn}=0.
\end{equation}
This expression should match the one-loop beta functions of the (unknown) dual CFT. This removes 15 real degrees of freedom and we can use the $\SU4$ action to remove another 15 real degrees of freedom, leaving 20 complex parameters, in agreement with the counting given by Kol~\cite{Kol02}. Recall that $\text{H}^{3}(\text{S}^{7})=0$ and so there are no marginal deformations due to closed $(1,2)$-forms $\chi$.

The $\beta$-deformed $\text{S}^{7}$ solution was first given in \cite{LM05}, which we reproduce in appendix \ref{sec:deform_appendix}. Taking $f=2\ii\gamma z_{1}z_{2}z_{3}z_{4}$, where $\gamma$ is real, and using our frame for $\text{S}^{7}$, one can check that our expression (\ref{eq:flux}) reproduces the four-form flux of the first-order $\beta$-deformed $\text{S}^{7}$ solution. Notice that we can also take $f\propto\gamma z_{1}z_{2}z_{3}z_{4}$, where we have dropped a factor of $\ii$ compared with the LM solution. This will also solve the moment map conditions and is a different marginal deformation, similar to the full complex $\beta$-deformation of $\mathcal{N}=4$ super Yang–Mills

\subsection{$\text{Q}^{1,1,1}$}

As an AdS$_{4}$ background in M-theory, the Sasaki–Einstein manifold Q$^{1,1,1}$ preserves eight supercharges. The dual field theory was identified in \cite{Benini:2009qs} and studied further in \cite{Cheon:2011th,JKPS11}. Viewing Q$^{1,1,1}$ as a $\Uni{1}$ fibration over $\mathbb{C}\mathbb{P}^{1}\times\mathbb{C}\mathbb{P}^{1}\times\mathbb{C}\mathbb{P}^{1}$, the metric\footnote{The metric has been scaled to ensure $R_{\mu\nu}=6g_{\mu\nu}$.} can be written as~\cite{DFN84,PP84}
\begin{equation}
\dd s^{2}(\text{Q}^{1,1,1})=\tfrac{1}{16}\biggl(\dd\psi+\sum_{i=1}^{3}\cos\theta_{i}\,\dd\phi_{i}\biggr)^{2}+\tfrac{1}{8}\sum_{i=1}^{3}(\dd\theta_{i}^{2}+\sin^{2}\theta_{i}\,\dd\phi_{i}^{2}).\label{eq:Q111_metric}
\end{equation}
We introduce an explicit frame in terms of the coordinates on $\text{Q}^{1,1,1}$:
\begin{equation}
\begin{split}e^{1}+\ii\,e^{2} & =\tfrac{1}{2\sqrt{2}}\ee^{\ii\psi/3}(\ii\,\dd\theta_{1}+\sin\theta_{1}\,\dd\phi_{1}),\\
e^{3}+\ii\,e^{4} & =\tfrac{1}{2\sqrt{2}}\ee^{\ii\psi/3}(\ii\,\dd\theta_{2}+\sin\theta_{2}\,\dd\phi_{2}),\\
e^{5}+\ii\,e^{6} & =\tfrac{1}{2\sqrt{2}}\ee^{\ii\psi/3}(\ii\,\dd\theta_{3}+\sin\theta_{3}\,\dd\phi_{3}),\\
e^{7} & =\tfrac{1}{4}(\dd\psi+\cos\theta_{1}\,\dd\phi_{1}+\cos\theta_{2}\,\dd\phi_{2}+\cos\theta_{3}\,\dd\phi_{3}).
\end{split}
\end{equation}
Using this frame, one can check that the complex, symplectic and contact structures given in (\ref{eq:SE_frame}) satisfy the algebraic and differential conditions (\ref{eq:SE_algebraic}) and (\ref{eq:SE_diff}).

Up to closed three-forms, the deformation is parametrised by a holomorphic function $f$ that descends from the Calabi–Yau cone over $\text{Q}^{1,1,1}$. The deformations are marginal if $f$ is of weight four under the Reeb vector. In our parametrisation, the Reeb vector is
\begin{equation}
\xi=4\partial_{\psi}.
\end{equation}

The cone over Q$^{1,1,1}$ is described by an embedding in $\mathbb{C}^{8}$ using eight complex coordinates $w_{i}$ that satisfy nine constraint equations. The explicit form of the coordinates is~\cite{BRS10}
\begin{equation}
\begin{aligned}w_{1} & =\ee^{\frac{\ii}{2}(\psi+\phi_{1}+\phi_{2}+\phi_{3})}c_{\theta_{1}/2}c_{\theta_{2}/2}c_{\theta_{3}/2}, & w_{2} & =\ee^{\frac{\ii}{2}(\psi-\phi_{1}-\phi_{2}-\phi_{3})}s_{\theta_{1}/2}s_{\theta_{2}/2}s_{\theta_{3}/2},\\
w_{3} & =\ee^{\frac{\ii}{2}(\psi+\phi_{1}-\phi_{2}-\phi_{3})}c_{\theta_{1}/2}s_{\theta_{2}/2}s_{\theta_{3}/2}, & w_{4} & =\ee^{\frac{\ii}{2}(\psi-\phi_{1}+\phi_{2}+\phi_{3})}s_{\theta_{1}/2}c_{\theta_{2}/2}c_{\theta_{3}/2},\\
w_{5} & =\ee^{\frac{\ii}{2}(\psi+\phi_{1}+\phi_{2}-\phi_{3})}c_{\theta_{1}/2}c_{\theta_{2}/2}s_{\theta_{3}/2}, & w_{6} & =\ee^{\frac{\ii}{2}(\psi-\phi_{1}+\phi_{2}-\phi_{3})}s_{\theta_{1}/2}c_{\theta_{2}/2}s_{\theta_{3}/2},\\
w_{7} & =\ee^{\frac{\ii}{2}(\psi+\phi_{1}-\phi_{2}+\phi_{3})}c_{\theta_{1}/2}s_{\theta_{2}/2}c_{\theta_{3}/2}, & w_{8} & =\ee^{\frac{\ii}{2}(\psi-\phi_{1}-\phi_{2}+\phi_{3})}s_{\theta_{1}/2}s_{\theta_{2}/2}c_{\theta_{3}/2}.
\end{aligned}
\end{equation}
The embedding coordinates $w_{i}$ are charge $+2$ under the Reeb vector field, so the general form of the function $f$ is
\begin{equation}
f=f^{ij}w_{i}w_{j},
\end{equation}
where $f^{ij}$ is symmetric with complex entries. There are generically 36 complex degrees of freedom in such a symmetric rank-two tensor, but 9 of them will not contribute to $f$ due to the constraints on the $w_{i}$. Thus there are 27 complex degrees of freedom corresponding to 27 marginal deformations. We can also use homogeneous coordinates $A_{a}$, $B_{\dot{a}}$ and $C_{\ddot{a}}$ that are related to the $w_{i}$ by~\cite{FKR09}
\begin{equation}
\begin{aligned}w_{1} & =A_{1}B_{2}C_{1}, & w_{2} & =A_{2}B_{1}C_{2}, & w_{3} & =A_{1}B_{1}C_{2}, & w_{4} & =A_{2}B_{2}C_{1},\\
w_{5} & =A_{1}B_{1}C_{1}, & w_{6} & =A_{2}B_{1}C_{1}, & w_{7} & =A_{1}B_{2}C_{2}, & w_{8} & =A_{2}B_{2}C_{2}.
\end{aligned}
\end{equation}
 We can then write the generic deformation as
\begin{equation}
f=f^{ab,\dot{a}\dot{b},\ddot{a}\ddot{b}}A_{a}B_{\dot{a}}C_{\ddot{a}}A_{b}B_{\dot{b}}C_{\ddot{b}},
\end{equation}
where $f^{ab,\dot{a}\dot{b},\ddot{a}\ddot{b}}$ is symmetric in $(ab)$, $(b\dot{b})$ and $(\ddot{a}\ddot{b})$. 

We can think of Q$^{1,1,1}$ as a $\Uni{1}$ fibration over a $\mathbb{C}\mathbb{P}^{1}\times\mathbb{C}\mathbb{P}^{1}\times\mathbb{C}\mathbb{P}^{1}$ base, so there is an $\SU{2}^{3}$ isometry that leaves the solution invariant, giving an $\SU{2}^{3}$'s worth of fixed-point symmetries. Again, we want to take a symplectic reduction of the space of couplings by the action of $\SU{2}^{3}$. The moment map for the first $\SU{2}$ action is
\begin{equation}
\mu_{\SU{2}}=f^{ac,\dot{a}\dot{b},\ddot{a}\ddot{b}}\bar{f}_{bc,\dot{a}\dot{b},\ddot{a}\ddot{b}}-\tfrac{1}{2}\delta_{\phantom{a}b}^{a}f^{cd,\dot{a}\dot{b},\ddot{a}\ddot{b}}\bar{f}_{cd,\dot{a}\dot{b},\ddot{a}\ddot{b}},
\end{equation}
and the others follow by swapping undotted for dotted or double-dotted indices. The conformal manifold of exactly marginal deformations that preserve eight supercharges is given by the symplectic reduction
\begin{equation}
\mathcal{M}_{\text{c}}=\{f^{ab,\dot{a}\dot{b},\ddot{a}\ddot{b}}\}\qquotient\SU{2}^{3}.
\end{equation}
The three moment maps for $\SU{2}$ gives 9 real conditions on the $f^{ab,\dot{a}\dot{b},\ddot{a}\ddot{b}}$, and we can remove another 9 degrees of freedom using $\SU{2}^{3}$ rotations of the couplings. In addition, $\text{H}^{3}(\text{Q}^{1,1,1})=0$ and so there are no marginal deformations due to closed $(1,2)$-forms $\chi$. Thus, the conformal manifold is $27-9=18$ complex dimensional.

The $\beta$-deformed $\text{Q}^{1,1,1}$ solution was first given in \cite{AV05,GLMW05}, which we reproduce in appendix \ref{sec:deform_appendix}. Taking $f\propto\gamma w_{1}w_{2}=\gamma A_{1}A_{2}B_{1}B_{2}C_{1}C_{2}$, where $\gamma$ is real, and using our frame for $\text{Q}^{1,1,1}$, one can check that our expression (\ref{eq:flux}) reproduces the four-form flux of the first-order $\beta$-deformed solution.

\subsection{$\text{M}^{1,1,1}$}

As an AdS$_{4}$ background in M-theory, the Sasaki–Einstein manifold M$^{1,1,1}$ preserves eight supercharges. Following the presentation in~\cite{PZ09}, the metric on M$^{1,1,1}$ can be written as
\begin{equation}
\begin{split}\dd s^{2}(\text{M}^{1,1,1}) & =\tfrac{3}{4}\Bigl(\dd\mu^{2}+\tfrac{1}{4}s_{\mu}^{2}c_{\mu}^{2}(\dd\psi+c_{\tilde{\theta}}\,\dd\tilde{\phi})^{2}+\tfrac{1}{4}s_{\mu}^{2}(\dd\tilde{\theta}^{2}+s_{\tilde{\theta}}^{2}\,\dd\tilde{\phi}^{2})\Bigr)\\
 & \eqspace+\tfrac{1}{8}(\dd\theta^{2}+s_{\theta}^{2}\,\dd\phi^{2})+\tfrac{1}{64}(\dd\tau+\lambda+2c_{\theta}\,\dd\phi)^{2},
\end{split}
\label{eq:M111_metric}
\end{equation}
where $\lambda=\tfrac{1}{2}(1+3\cos2\mu)\dd\psi-3\cos\tilde{\theta}\sin^{2}\mu\dd\tilde{\phi}$.\footnote{Note that the $\lambda$ we use differs from that of \cite{AV05,PZ09} by $2\dd\psi$.} We can introduce an explicit frame in terms of the coordinates on $\text{M}^{1,1,1}$:
\begin{equation}
\begin{split}e^{1}+\ii\,e^{2} & =\tfrac{\sqrt{3}}{2}\ee^{\ii\tau/6}\Big(\dd\mu-\tfrac{1}{4}\ii\sin2\mu(\dd\psi+\cos\tilde{\theta}\,\dd\tilde{\phi})\Big),\\
e^{3}+\ii\,e^{4} & =\tfrac{\sqrt{3}}{4}\ee^{\ii\tau/6}\sin\mu(\dd\tilde{\theta}+\ii\sin\tilde{\theta}\,\dd\tilde{\phi}),\\
e^{5}+\ii\,e^{6} & =\tfrac{1}{2\sqrt{2}}\ee^{\ii\tau/6}(\dd\theta-\ii\sin\theta\,\dd\phi),\\
e^{7} & =\tfrac{1}{8}(\dd\tau+\lambda+2\cos\theta\,\dd\phi).
\end{split}
\end{equation}
Using this frame, one can check that the complex, symplectic and contact structures given in (\ref{eq:SE_frame}) satisfy the algebraic and differential conditions (\ref{eq:SE_algebraic}) and (\ref{eq:SE_diff}).

Up to closed three-forms, the deformation is parametrised by a holomorphic function $f$ that descends from the Calabi–Yau cone over $\text{M}^{1,1,1}$. The deformations are marginal if $f$ is of weight four under the Reeb vector. In our parametrisation, the Reeb vector is
\begin{equation}
\xi=8\partial_{\tau}.
\end{equation}
The cone over M$^{1,1,1}$ can be described by an embedding in $\mathbb{C}^{30}$~\cite{FFGRTZZ00}. Instead we use homogeneous coordinates $U_{i}$ and $V_{a}$ which are charge $+8/9$ and $+2/3$ respectively under the Reeb vector field~\cite{GLMW05}, so the general form of the function $f$ is
\begin{equation}
f=f^{ijk,ab}U_{i}U_{j}U_{k}V_{a}V_{b},
\end{equation}
where $f^{ijk,ab}$ is symmetric on $(ijk)$ and $(ab)$ with complex entries, transforming in the $(\rep{10},\rep{3})$ of $\SU3\times\SU2$. There are generically 30 complex degrees of freedom in such a tensor, thus there are 30 complex degrees of freedom corresponding to 30 marginal deformations.

Again we must consider if there are isometries at the $\text{M}^{1,1,1}$ point in the moduli space of couplings. M$^{1,1,1}$ is a $\Uni{1}$ fibration over a $\mathbb{C}\mathbb{P}^{2}\times\mathbb{C}\mathbb{P}^{1}$ base, so there is an $\SU{3}\times\SU2$ isometry that acts on the base, leaving the solution invariant. We can construct a moment map for the $\SU3\times\SU2$ action on the space of couplings and perform a symplectic reduction. The moment maps are
\begin{equation}
\begin{split}\mu_{\SU3} & =f^{ikl,ab}\bar{f}_{jkl,ab}-\tfrac{1}{3}\delta_{\phantom{i}j}^{i}f^{klm,ab}\bar{f}_{klm,ab},\\
\mu_{\SU2} & =f^{ijk,ac}\bar{f}_{ikl,bc}-\tfrac{1}{2}\delta_{\phantom{a}b}^{a}f^{ijk,cd}\bar{f}_{ijk,cd}.
\end{split}
\end{equation}
The conformal manifold of exactly marginal deformations that preserve eight supercharges is given by the symplectic reduction
\begin{equation}
\mathcal{M}_{\text{c}}=\{f^{ijk,ab}\}\qquotient\SU3\times\SU2.
\end{equation}
The moment maps give $8+3$ real conditions on the $f^{ijk,ab}$, and we can remove another 11 degrees of freedom using rotations of the couplings. In addition, $\text{H}^{3}(\text{M}^{1,1,1})=0$ and so all global three-forms are trivial in cohomology~\cite{FFGRTZZ00}. This means there are no marginal deformations due to closed $(1,2)$-forms $\chi$. Thus, the conformal manifold is $30-11=19$ complex dimensional.

The $\beta$-deformed $\text{M}^{1,1,1}$ solution was first given in \cite{AV05,GLMW05}, which we reproduce in appendix \ref{sec:deform_appendix}. Taking $f\propto i\gamma\ee^{\ii\tau/2}\sin\theta\sin\tilde{\theta}\sin^{2}\mu\cos\mu$, where $\gamma$ is real, and using our frame for $\text{M}^{1,1,1}$, one can check that our expression (\ref{eq:flux}) reproduces the four-form flux of the first-order $\beta$-deformed solution.

\section{Lunin–Maldacena from a tri-vector deformation\label{sec:The-Lunin=002013Maldacena-background}}

Up to now we have found the first-order deformations that are dual to the marginal deformations of the corresponding field theories. Ideally we would find the all-orders supergravity solutions dual to the deformed field theories. As evidenced by many years of work on this topic, this is no easy feat~\cite{HT09,MPZ06,AGGPW16,AKY02,Dlamini:2016aaa,Mansson:2008xv}. A general formalism for finding these backgrounds is currently out of reach, but we do have a class of solutions for which the full supergravity backgrounds are known: the Lunin–Maldacena solutions. These solutions were found for type IIB backgrounds with at least a $\Uni{1}^{2}$ isometry or M-theory backgrounds with a $\Uni{1}^{3}$ isometry~\cite{LM05}. The examples we have considered thus far are exactly of this sort, and so we might hope that the full supergravity backgrounds can be obtained by deforming the generalised structures we have given. We now show that this is the case: the LM solutions can be obtained by acting with a tri-vector deformation.

Let us focus on the $\text{S}^{7}$ background with H and V structures given by (\ref{eq:H_structure}) and (\ref{eq:V_structure}). We work in a gauge where the six-form potential is
\begin{equation}
\tilde{A}=\tfrac{1}{2}\dd\psi\wedge\sigma\wedge\omega\wedge\omega.
\end{equation}
For this section we will include the twisting by $\tilde{A}$ in the structures so that
\begin{equation}
J_{\alpha}\mapsto\ee^{\tilde{A}}J_{\alpha}\ee^{-\tilde{A}},\qquad X\mapsto\ee^{\tilde{A}}X.\label{eq:twisting_included}
\end{equation}

We take the LM deformation to be defined by an adjoint element given by
\begin{equation}
\alpha_{\text{LM}}=\gamma(\partial_{1}\wedge\partial_{2}\wedge\partial_{3}-\partial_{1}\wedge\partial_{2}\wedge\partial_{4}+\partial_{1}\wedge\partial_{3}\wedge\partial_{4}-\partial_{2}\wedge\partial_{3}\wedge\partial_{4}),\label{eq:LM_def}
\end{equation}
where we are using the shorthand $\partial_{i}=\partial_{\phi_{i}}$. Acting with this adjoint element on the twisted structures, one can check it leaves $X$ invariant
\begin{equation}
X^{\text{LM}}=\ee^{\alpha_{\text{LM}}}X=X,\label{eq:X_invariant}
\end{equation}
 and truncates at first order on $J_{3}$ and second order on $J_{+}$
\begin{equation}
\begin{split}J_{3}^{\text{LM}}=\ee^{\alpha_{\text{LM}}}J_{3}\ee^{-\alpha_{\text{LM}}} & =J_{3}+[\alpha_{\text{LM}},J_{3}],\\
J_{+}^{\text{LM}}=\ee^{\alpha_{\text{LM}}}J_{+}\ee^{-\alpha_{\text{LM}}} & =J_{+}+[\alpha_{\text{LM}},J_{+}]+\tfrac{1}{2}[\alpha_{\text{LM}},[\alpha_{\text{LM}},J_{+}]].
\end{split}
\end{equation}
One can check that the deformed structures are correctly normalised (with $\kappa^{2}$ unchanged from the $\text{S}^{7}$ background) and they are compatible. In fact, this must be the case as $\ee^{\alpha^{\text{LM}}}$ acts in $\Ex{7(7)}$ so it cannot change the invariant volume, nor the relation $J_{\alpha}\cdot X=0$. In the field theory, this means the central charge is not changed by the corresponding exactly marginal deformation.

The question is whether these new structures give an $\mathcal{N}=2$ solution. For this we need to check whether they satisfy the supersymmetry conditions given in (\ref{eq:moment_maps}), (\ref{eq:K_cond}) and (\ref{eq:u1_charge}). The second of these, equivalent to $\Dorf_{X}\bar{X}=0$, is trivially satisfied as the V structure is invariant under the deformation. The third condition is also satisfied. To see this note that the Dorfman derivative of a generalised tensor $R$ along a generalised vector $V$ is
\begin{equation}
\Dorf_{V}R=\mathcal{L}_{v}R-[\dd\omega+\dd\sigma,R],
\end{equation}
where $v$, $\omega$ and $\sigma$ are vector, two- and five-form components of $V$. The two- and five-form components of $X$ are closed and so (\ref{eq:u1_charge}) reduces to
\begin{equation}
\mathcal{L}_{\xi}J_{\alpha}^{\text{LM}}=\epsilon_{\alpha\beta\gamma}\lambda_{\beta}J_{\gamma}^{\text{LM}}.
\end{equation}
One can check that $\alpha_{\text{LM}}$ is uncharged under $\xi$ and so this condition is satisfied. 

We only need to check the moment maps. Using the form of the Dorfman derivative, the moment maps can be written as
\begin{equation}
\mu_{\alpha}(V)=-\tfrac{1}{2}\epsilon_{\alpha\beta\gamma}\int_{M}\tr(J_{\beta}^{\text{LM}}\mathcal{L}_{v}J_{\gamma}^{\text{LM}})-2\int_{M}\kappa\,\tr\bigl((\dd\omega+\dd\sigma)J_{\alpha}^{\text{LM}}\bigr)=\lambda_{\alpha}\int_{M}s(V,\hat{K}),
\end{equation}
where $\kappa^{2}$ and $\hat{K}$ are unchanged from the $\text{S}^{7}$ background.\footnote{One can see this by noting that $X^{\text{LM}}=X$ or that the deformation does not have an $\mathbb{R}^{+}$ part and so does not change the invariant volume.} Recall that this equation has to hold for any $V$ (and so for arbitrary choices of $v$, $\omega$ and $\sigma$). We denote the difference between the $\text{S}^{7}$ and LM structures as
\begin{equation}
J_{\alpha}^{\text{LM}}=J_{\alpha}+\delta J_{\alpha},
\end{equation}
and recall that the moment maps can be written as
\begin{equation}
\mu_{3}(V)\propto\int_{M}\tr(J_{3}\Dorf_{V}J_{+}),\qquad\mu_{+}(V)\propto\int_{M}\tr(J_{-}\Dorf_{V}J_{+}).
\end{equation}
With this, the moment map conditions simplify to
\begin{align}
\int_{M}\kappa\,\tr(\dd\omega+\dd\sigma,\delta J_{\alpha}) & =0.\label{eq:one}\\
\int_{M}\tr(J_{3}\mathcal{L}_{v}\delta J_{+})+\int_{M}\tr(\delta J_{3}\mathcal{L}_{v}J_{+})+\int_{M}\tr(\delta J_{3}\mathcal{L}_{v}\delta J_{+}) & =0,\label{eq:three}\\
\int_{M}\tr(J_{-}\mathcal{L}_{v}\delta J_{+})+\int_{M}\tr(\delta J_{-}\mathcal{L}_{v}J_{+})+\int_{M}\tr(\delta J_{-}\mathcal{L}_{v}\delta J_{+}) & =0.\label{eq:four}
\end{align}
At this point we need to know that $\delta J_{3}$ and $\delta J_{+}$ have the following non-zero components
\begin{equation}
\begin{split}\delta J_{3} & =\kappa(a_{3}+\alpha_{3}),\\
\delta J_{+} & =\kappa(l_{+}+r_{+}+\alpha_{+}+\tilde{\alpha}_{+}),
\end{split}
\end{equation}
where $l_{+}$ is a scalar, $r_{+}$ is a $\gl7$ element, $a_{3}$ is a three-form, $\alpha_{+}$ and $\alpha_{3}$ are three-vectors and $\tilde{\alpha}_{+}$ is a six-vector. Note that $\alpha_{+}$ is second order in the deformation, and the other components are first order. 

The first condition (\ref{eq:one}) simplifies to differential conditions on the three- and six-vector components of $\delta J_{\alpha}$, namely
\begin{equation}
\dd\star\alpha_{3}^{\flat}=0,\qquad\dd\star\alpha_{+}^{\flat}=0,\qquad\dd\star\tilde{\alpha}_{+}^{\flat}=0
\end{equation}
where $\star$ and $\flat$ (lowering the indices of a $p$-vector) are with respect to the $\text{S}^{7}$ metric. In other words, $\alpha_{3}^{\flat}$, $\alpha_{+}^{\flat}$ and $\tilde{\alpha}_{+}^{\flat}$ must be co-closed. One can check these are satisfied using the expressions for $J_{\alpha}$ in (\ref{eq:H_structure}), the expressions for the invariant forms in (\ref{eq:SE_frame}), the frame for $\text{S}^{7}$ in (\ref{eq:S7_frame}), and the deformation $\alpha_{\text{LM}}$ in (\ref{eq:LM_def}).\footnote{This is most easily checked using a Mathematica package such as Atlas2 or xAct.}

The remaining conditions are those that contain a Lie derivative along $v$. For the $\mu_{3}$ moment map, we have\footnote{Note that the $\Omega^{\sharp}\lrcorner\tilde{A}$ terms come from including the six-form potential in the structures, as shown in (\ref{eq:twisting_included}).}
\begin{equation}
\begin{split}\delta\mu_{3} & =\int_{M}\tr(J_{3}\mathcal{L}_{v}\delta J_{+})+\int_{M}\tr(\delta J_{3}\mathcal{L}_{v}J_{+})+\int_{M}\tr(\delta J_{3}\mathcal{L}_{v}\delta J_{+})\\
 & =\int_{M}\tr\bigl(2\ii\,\kappa J_{+}\mathcal{L}_{v}\alpha_{\text{LM}}+\kappa\,a_{3}\mathcal{L}_{v}(\kappa\alpha_{+})\bigr)\\
 & =\int_{M}\Bigl(\ii\kappa^{2}\imath_{v}\dd\bigl(\alpha_{\text{LM}}\lrcorner(\Omega-\Omega^{\sharp}\lrcorner\tilde{A})\bigr)+\ii(\Omega-\Omega^{\sharp}\lrcorner\tilde{A})\wedge\imath_{v}\dd\star\alpha_{\text{LM}}^{\flat}\\
 & \eqspace\phantom{\int_{M}\Bigl(}+\ii(\Omega-\Omega^{\sharp}\lrcorner\tilde{A})\wedge\dd\imath_{v}\star\alpha_{\text{LM}}^{\flat}-\imath_{v}\dd a_{3}\wedge\star\alpha_{+}^{\flat}+\imath_{v}a_{3}\wedge\dd\star\alpha_{+}^{\flat})\Bigr)\\
 & =\int_{M}\Bigl(\ii\kappa^{2}\imath_{v}\dd\bigl(\alpha_{\text{LM}}\lrcorner(\Omega-\Omega^{\sharp}\lrcorner\tilde{A})\bigr)+\ii\dd(\Omega-\Omega^{\sharp}\lrcorner\tilde{A})\wedge\imath_{v}\star\alpha_{\text{LM}}^{\flat}\Bigr)\\
 & =0,
\end{split}
\end{equation}
where we have used $\alpha_{+}\lrcorner a_{3}=0$, $\dd a_{3}=0$, $\dd\star\alpha_{+}^{\flat}=0$ and $\dd\star\alpha_{\text{LM}}^{\flat}=0$ to reach the penultimate line. The remaining terms cancel against each other to give zero. Finally, the $\mu_{+}$ moment map gives
\begin{equation}
\begin{split}\delta\mu_{+} & =\int_{M}\tr(J_{-}\mathcal{L}_{v}\delta J_{+})+\int_{M}\tr(\delta J_{-}\mathcal{L}_{v}J_{+})+\int_{M}\tr(\delta J_{-}\mathcal{L}_{v}\delta J_{+})\\
 & =\int_{M}\tr\bigl(-\tfrac{1}{2}[J_{-},J_{+}][\mathcal{L}_{v}\alpha_{\text{LM}},\alpha_{\text{LM}}]\bigr)\\
 & =\int_{M}\tr(-2\ii\kappa^{2}\,a_{3}\,\mathcal{L}_{v}\alpha_{\text{LM}})\\
 & =\int_{M}2\ii\,\mathcal{L}_{v}a_{3}\wedge\star\alpha_{\text{LM}}^{\flat}\\
 & =\int_{M}\bigl(2\ii\,\imath_{v}\dd a_{3}\wedge\star\alpha_{\text{LM}}^{\flat}-2\ii\,\imath_{v}a_{3}\wedge\dd\star\alpha_{\text{LM}}^{\flat}\bigr)\\
 & =0,
\end{split}
\end{equation}
 where we have used $\alpha_{\text{LM}}\lrcorner a_{3}=0$, $\dd a_{3}=0$, and $\dd\star\alpha_{\text{LM}}^{\flat}=0$ to reach the final line. With this we see the deformed structures satisfy all the conditions to define an $\mathcal{N}=2$ background.

We now have a non-trivial deformation of the $\text{AdS}_{4}\times\text{S}^{7}$ solution that preserves $\mathcal{N}=2$ supersymmetry to all orders in the deformation parameter. It might be slightly mysterious how we did this. Looking at the form of the metric for the LM deformed solution, given in appendix \ref{sec:deform_appendix}, we see it depends on $(1+\gamma^{2}\Sigma)^{-1}$ where $\Sigma$ is a function of the coordinates on S$^{7}$. In particular, upon expanding as a power series in $\gamma$, we see it is corrected to all orders. The objects defining the ESE structure, $\{J_{\alpha},K\}$, are not corrected to all orders: $J_{\alpha}$ is corrected to second order and $K$ is actually unchanged. Where is the extra non-linearity hiding? The answer is in how one goes from $\{J_{\alpha},K\}$ to the supergravity fields.\footnote{This is also the case for the analogous calculation in type II on $\text{AdS}_{5}\times\text{S}^{5}$. The pure spinors are corrected to second order and the non-linearity appears in passing from these to the generalised metric~\cite{MPZ06,HT09}.}

The supergravity fields $\{\Delta,g,A,\tilde{A}\}$ are encoded in the generalised metric~\cite{CSW11,CSW14}. The generalised metric is equivalent to a generalised $\SU8$ structure. The ESE structure (defined by $\{J_{\alpha},K\}$) is a generalised $\SU6$ structure, and so it also defines an $\SU8$ structure. In other words, $\{J_{\alpha},K\}$ defines a generalised metric and thus the supergravity fields of the solutions. In general, the relation between these objects will be complicated and non-linear, and has not been known till now.

\section{Generalised metric\label{sec:Generalised-metric}}

In this section we outline how the generalised $\SU6$ structure that characterises a supersymmetric flux background specifies a generalised metric (the data of a metric on $M$, a warp factor and three- and six-form potentials). We give the formula for the generalised metric in terms of $\{J_{\alpha},K,\hat{K}\}$. 

Let us begin by reviewing how this works for a conventional $G$-structure. The $G$-structures that characterise supersymmetric backgrounds without flux all define a Riemannian metrics as the reduced structure group $G$ is a subgroup of $\Orth{d}$. In general, the relation between the invariant tensors that characterise the $G$-structure and the metric is complicated. Let us look at a couple of examples.

\subsubsection*{SU(3) structure}

In six dimensions an $\SU3$ structure is defined by a real two-form $\omega$ and a complex three-form $\Omega$. The three-form defines a complex structure $I$ via~\cite{Hitchin00}
\begin{equation}
I=\pm\frac{\tilde{I}}{H(\rho)},
\end{equation}
where $\tilde{I}$, $\rho$ and $H$ are defined by
\begin{equation}
\tilde{I}=\pm\epsilon^{ik_{1}\ldots k_{5}}\rho_{jk_{1}k_{2}}\rho_{k_{3}k_{4}k_{5}},\qquad\rho=\re\Omega,\qquad H(\rho)=\sqrt{-\tfrac{1}{6}\tr\tilde{I}^{2}}.
\end{equation}
The metric is then defined by
\begin{equation}
g(u,v)=\omega(u,Iv).
\end{equation}
Though the expression for $g$ is simple in terms of $\omega$ and $I$, the relation between $I$ and $\Omega$ is non-linear.

\subsubsection*{\texorpdfstring{$\Gx{2}$}{G_2} structure}

In seven dimensions a $\Gx2$ structure is defined by a three-form $\phi$ and a four-form $\psi=\star\phi$. Again there is an expression for the metric in term of the invariant forms:
\begin{equation}
g_{ab}=(\det s)^{-1/9}s_{ab},
\end{equation}
where $s_{ab}$ is given by
\begin{equation}
s_{ab}=-\tfrac{1}{144}\phi_{ac_{1}c_{2}}\phi_{bc_{3}c_{4}}\phi_{c_{5}c_{6}c_{7}}\epsilon^{c_{1}\ldots c_{7}}.
\end{equation}
In this case the metric is clearly a non-linear function of the invariant form $\phi$.

\subsubsection*{Generalised SU(6) structure}

The analogue of $\Orth{d}$ for $\Ex{7(7)}\times\mathbb{R}^{+}$ generalised geometry is $\SU8$. A reduction of the structure group to $\SU8$ signals the existence of a generalised metric that encodes the bosonic fields of the supergravity solution, namely the metric, warp factor, and three- and six-form potentials. The Sasaki–Einstein backgrounds and the deformations that we have discussed in this paper all define generalised $\SU6$ structures. As $\SU6\subset\SU8$ they automatically define a generalised metric. As with the case of $\Gx2$ structures, the relation between the $\SU6$ invariant objects $\{J_{\alpha},K,\hat{K}\}$ and the generalised metric $G$ is expected to be non-linear. We now give an expression for this generalised metric so that given $\{J_{\alpha},K,\hat{K}\}$ one can reconstruct the supergravity fields.

First we need an expression for the quartic invariant $q$ of $\Ex{7(7)}$. Using the requirement that $q$ is invariant under $\Ex{7(7)}$ transformations, we can fix the quartic invariant (up to scale) to
\begin{equation}
\begin{split}q(V) & =-\tfrac{1}{4}(v\lrcorner\omega)\wedge\sigma\wedge t\,\widetilde{\vol}_{7}+\tfrac{1}{8}(v\lrcorner t)\,(\omega\wedge\sigma)\widetilde{\vol}_{7}-\tfrac{1}{24}\omega\wedge\omega\wedge\omega\wedge t\,\widetilde{\vol}_{7}\\
 & \eqspace+\tfrac{1}{48}(\sigma\wedge e^{m_{1}}\wedge e^{m_{2}})\,(\imath_{\hat{e}_{m_{1}}}\imath_{\hat{e}_{m_{2}}}\imath_{v}\sigma)\wedge\sigma-\tfrac{1}{16}(v\lrcorner t)^{2}\,\widetilde{\vol}_{7}^{2}\\
 & \eqspace+\tfrac{1}{16}(\omega\wedge\sigma)^{2}-\tfrac{1}{16}(e^{m_{1}}\wedge(\imath_{\hat{e}_{m_{2}}}\omega)\wedge\sigma)\,(e^{m_{2}}\wedge(\imath_{\hat{e}_{m_{1}}}\omega)\wedge\sigma),
\end{split}
\end{equation}
where $q(V)\equiv q(V,V,V,V)$. Note that we have written the one-form valued in $\ext^{7}T^{*}M$ as $\tau=t\otimes\widetilde{\vol}_{7},$ where $t$ is a one-form and $\widetilde{\vol}_{7}$ is any non-vanishing seven-form on $M$ – $q(V)$ depends on $\tau$ and not the particular factorisation into $t$ and $\widetilde{\vol}_{7}$.\footnote{Alternatively, one can express this in terms of $\tau$ alone using the ``$j$ notation'' of \cite{CSW11}.} For the Sasaki–Einstein structures given in section 2, one can show
\begin{equation}
q(K)=\tfrac{1}{4}\vol_{7}^{2},\qquad q(\hat{K})=\tfrac{1}{4}\vol_{7}^{2},\qquad s(K,\hat{K})=\vol_{7},
\end{equation}
so that we have $2\sqrt{q(K)}=s(K,\hat{K})$ in agreement with equation (\ref{eq:moment_maps}).

We now construct the generalised metric from $q$, $\hat{J}_{\alpha}$, $K$ and $\hat{K}$. Here $\hat{J}_{\alpha}=\kappa^{-1}J_{\alpha}$ are the unweighted structures which do not have a factor of $\kappa$. The generalised metric is given by\footnote{The form of this expression was inspired by discussions with D.~Waldram and M.~Petrini on the analogous expression for $\Ex{6(6)}$.}
\begin{equation}
\begin{split}\tfrac{1}{24}G(V,V) & =-\frac{q(K,K,V,V)}{q(K)}+\frac{q(K,K,K,V)^{2}}{q(K)^{2}}+\tfrac{1}{2}\frac{q(\hat{K},\hat{K},V,V)}{q(K)}\\
 & \eqspace+\frac{q(K,K,\hat{J}_{3}\cdot V,\hat{J}_{3}\cdot V)}{q(K)}.
\end{split}
\end{equation}
Note that this expression is not unique – there are other terms that one could construct but they are not independent of those already included. One can check that this reproduces the standard metric $g=\sum_{a=1}^{7}e^{a}\otimes e^{a}$ when the expressions for $K$, $\hat{K}$ and $J_{\alpha}$ given in section 2 are used. 

\subsubsection*{Lunin–Maldacena solution}

We now want an expression that relates the generalised metric for the $\text{S}^{7}$ solution to the deformed LM solution. This will allow us to check that the deformation given in section 5 reproduces the metric, warp factor and fluxes of the LM solution.

The generalised metric for the LM solution is given by taking $J_{\alpha}\rightarrow\ee^{\alpha_{\text{LM}}}J_{\alpha}\ee^{-\alpha_{\text{LM}}}$ where $J_{\alpha}$ is the S$^{7}$ structure (and we recall from (\ref{eq:X_invariant}), $K$ and $\hat{K}$ are invariant). Using that $q$ is invariant under $\Ex{7(7)}$ transformations and that $\ee^{\alpha_{\text{LM}}}$ acts in $\Ex{7(7)}$ alone, it is easy to show
\begin{equation}
G_{\text{LM}}(V,V)=G_{\text{S}^{7}}(\ee^{-\alpha_{\text{LM}}}V,\ee^{-\alpha_{\text{LM}}}V).
\end{equation}
Following \cite{LSW14}, the right-hand side can be written as
\begin{equation}
G_{\text{S}^{7}}(\ee^{-\alpha_{\text{LM}}}V,\ee^{-\alpha_{\text{LM}}}V)=|\ee^{-\tilde{A}}\ee^{-\alpha_{\text{LM}}}V|_{\text{S}^{7}}^{2},
\end{equation}
where $\tilde{A}$ is the six-form potential for S$^{7}$ and $|\cdot|_{\text{S}^{7}}^{2}$ is the norm with respect to the $\text{S}^{7}$ metric. Explicitly, for a generalised vector $\tilde{V}$ this is given by
\begin{equation}
\begin{split}|\tilde{V}|_{\text{S}^{7}}^{2} & =g_{mn}\tilde{v}^{m}\tilde{v}^{n}+\tfrac{1}{2!}g^{m_{1}n_{1}}g^{m_{2}n_{2}}\tilde{\omega}_{m_{1}m_{2}}\tilde{\omega}_{n_{1}n_{2}}+\tfrac{1}{5!}g^{m_{1}n_{1}}\ldots g^{m_{5}n_{5}}\tilde{\sigma}_{m_{1}\ldots m_{5}}\tilde{\sigma}_{n_{1}\ldots n_{5}}\\
 & \eqspace+\tfrac{1}{7!}g^{m_{1}n_{1}}\ldots g^{m_{8}n_{8}}\tilde{\tau}_{m_{1},m_{2}\ldots m_{8}}\tilde{\tau}_{n_{1},n_{2}\ldots n_{8}},
\end{split}
\end{equation}
where $g_{mn}$ is the metric on S$^{7}$ and $\tilde{v}$, $\tilde{\omega}$, etc.~are the components of $\tilde{V}$. Playing the same trick with $G_{\text{LM}}$ gives
\begin{equation}
G_{\text{LM}}(V,V)=|\ee^{-\Delta_{\text{LM}}}\ee^{-A_{\text{LM}}-\tilde{A}_{\text{LM}}}V|_{\text{LM}}^{2},
\end{equation}
where the norm is now with respect to the LM metric. We can now calculate $G_{\text{LM}}$ is terms of the metric and six-form potential for S$^{7}$:
\begin{equation}
|\ee^{-\Delta_{\text{LM}}}\ee^{-A_{\text{LM}}-\tilde{A}_{\text{LM}}}V|_{\text{LM}}^{2}=|\ee^{-\tilde{A}}\ee^{-\alpha_{\text{LM}}}V|_{\text{S}^{7}}^{2}.
\end{equation}
The right-hand side is completely determined by the metric $g$ and six-form potential $\tilde{A}$ for the S$^{7}$ solution – one needs to expand out both sides and compare the coefficients of the various components of $V$ to read off $\{\Delta_{\text{LM}},A_{\text{LM}},\tilde{A}_{\text{LM}}\}$, the warp factor, three-form and six-form potential for the deformed LM solution. We have checked that this reproduces the LM solution, given in appendix A.

\section{Conclusions}

We studied marginal deformations of $\mathcal{N}=2$ $\AdS4$ solutions with a Sasaki–Einstein internal space in eleven-dimensional supergravity using generalised geometry. We found that the first-order correction to these solutions is a four-form flux on the internal space. By viewing the perturbation as a deformation of a generalised structure, we were able to derive a general expression for the four-form flux in terms of a holomorphic function of charge $+4$ under the Reeb vector. We then discussed the explicit examples of S$^{7}$, Q$^{1,1,1}$ and M$^{1,1,1}$. Using an obstruction analysis, we found the conditions for the first-order deformations to extend all orders, thus identifying which marginal deformations are exactly marginal. Focussing on $\text{AdS}_{4}\times\text{S}^{7}$, we showed how the all-orders solution of Lunin and Maldacena can be encoded as a tri-vector deformation of a generalised structure. We also discussed how the supergravity fields can be recovered from the generalised structure and outlined how to do this for the Lunin–Maldacena solution.

Though we focussed on Sasaki–Einstein solutions, our approach is valid for any $\mathcal{N}=2$ AdS$_{4}$ solution in M-theory. The details, such as the invariant tensors that define the generalised structure, will change, but the method and the obstruction analysis holds for generic backgrounds. We hope this will prove useful for understanding marginal deformations of CFTs dual to more complicated flux backgrounds.

We saw how the all-orders Lunin–Maldacena solution could be described by a tri-vector deformation that truncates at second order. It would be interesting to see whether a similar approach can be used for the other deformations we found. These are the analogue of the Leigh–Strassler deformation of $\mathcal{N}=4$ super Yang–Mills, though there are many more of them. If one can find the all-orders supergravity backgrounds dual to these deformations, we would have a large number of new flux backgrounds. This would enable new non-trivial checks of the AdS/CFT correspondence and allow us to compute the metric on the conformal manifold. We hope to make progress on this in the near future.

\acknowledgments

We thank M.~Gra\~na and E.~Tasker for helpful discussions, and M.~Petrini and D.~Waldram for useful discussions, collaboration in the initial stages of this work and comments on the manuscript. AA is supported by a Junior Research Fellowship from Merton College, Oxford.

\appendix

\section{\texorpdfstring{$\gamma$}{Gamma}-deformed solutions\label{sec:deform_appendix}}

Here we summarise the results of the solution-generating technique of Lunin and Maldacena applied to $\AdS4$ solutions in M-theory~\cite{LM05}. We follow the general prescription laid out in~\cite{GLMW05}. The undeformed metric and four-form flux are
\begin{equation}
\dd s_{11}^{2}=\tfrac{1}{4}\dd s^{2}(\text{AdS}_{4})+\dd s^{2}(M),\qqq F_{4}=\tfrac{3}{8}\vol_{\text{AdS}}.
\end{equation}
Note that we have normalised the metric on the internal space $M$ to give $R_{mn}=6\,g_{mn}$.

First, we split the metric on $M$ into a three-torus and a four-dimensional space $M_{4}$
\begin{equation}
\dd s^{2}(M)=\dd s^{2}(\text{T}^{3})+\dd s^{2}(M_{4}).
\end{equation}
The metric on the torus is then expressed as\footnote{Note that we have renamed $\Delta$ in \cite{LM05,GLMW05} to $\Sigma$ to avoid confusion with the warp factor. }
\begin{equation}
\dd s^{2}(\text{T}^{3})=\Sigma^{1/3}M_{ab}\,D\varphi_{a}\,D\varphi_{b},
\end{equation}
where $D\varphi_{a}=\dd\varphi_{a}+A_{a}$, $\det M_{ab}=1$, and the one-forms $A_{a}$ depend on the undeformed metric. The eleven-dimensional solution obtained from the solution-generating technique is
\begin{equation}
\begin{split}\dd s_{11}^{2} & =G^{-1/3}\Bigl(\tfrac{1}{4}\dd s^{2}(\text{AdS}_{4})+\dd s^{2}(M_{4})+G\,\dd s^{2}(\text{T}^{3})\Bigr),\\
F & =\tfrac{3}{8}\vol_{\text{AdS}}-6\,\gamma\Sigma^{1/2}\vol_{4}-\gamma\,\dd(G\Sigma\,D\varphi_{1}\wedge D\varphi_{2}\wedge D\varphi_{3}),
\end{split}
\end{equation}
where $G=(1+\gamma^{2}\Sigma)^{-1}$ and $\vol_{4}$ is the volume form of $\dd s^{2}(M_{4})$. From this we see the contribution to the internal flux is
\begin{equation}
F=-6\,\gamma\Sigma^{1/2}\vol_{4}-\gamma\,\dd(\Sigma\,D\varphi_{1}\wedge D\varphi_{2}\wedge D\varphi_{3})+\mathcal{O}(\gamma^{2})+\ldots
\end{equation}
It is this expression that our first-order deformations reproduce. At this point the expression is completely general – to find the explicit form for a given background, we need to specify $\varphi_{a}$, $\Sigma$, $\vol_{4}$ and $A_{a}$ in terms of the coordinates for $\dd s^{2}(M)$. We now give these for the examples we discuss in the main text.

\subsubsection*{\texorpdfstring{S$^7$}{S7}\label{par:S7_def}}

The solution for S$^{7}$ was first given in \cite{LM05}. The angles parametrising the three-torus are
\begin{equation}
\varphi_{1}=3\psi-\phi_{1}-\phi_{2}-\phi_{3},\quad\varphi_{2}=2\psi-\phi_{1}-\phi_{2},\quad\varphi_{3}=\phi_{1}-\psi.
\end{equation}
The quantities $\Sigma$, $\vol_{4}$ and the $A_{a}$ are\footnote{Note that this corrects a typographical error in \cite{LM05}, where the term in the four-form flux coming from $\Sigma^{1/2}\vol_{4}$ was written with $s_{2\alpha}^{2}$ instead of $s_{2\alpha}s_{\alpha}^{2}$.}

\begin{equation}
\begin{split}\Sigma & =s_{\theta}^{4}s_{\alpha}^{2}\bigl(c_{\theta}^{2}c_{\alpha}^{2}+s_{\alpha}^{2}s_{\beta}^{2}c_{\beta}^{2}(c_{\theta}^{2}+s_{\theta}^{2}c_{\alpha}^{2})\bigr),\\
\vol_{4} & =-\Sigma^{-1/2}s_{\theta}^{5}c_{\theta}s_{2\alpha}s_{\alpha}^{2}s_{2\beta}\,\dd\theta\wedge\dd\alpha\wedge\dd\beta\wedge\dd\psi,\\
A_{1} & =\frac{-4(1+2\,c_{2\beta})c_{\theta}^{2}c_{\alpha}^{2}+s_{\alpha}^{2}s_{2\beta}^{2}(c_{\theta}^{2}+s_{\theta}^{2}c_{\alpha}^{2})}{4\,c_{\theta}^{2}c_{\alpha}^{2}+s_{\alpha}^{2}s_{2\beta}^{2}(c_{\theta}^{2}+s_{\theta}^{2}c_{\alpha}^{2})}\dd\psi,\\
A_{2} & =2\frac{-4\,c_{\theta}^{2}c_{\alpha}^{2}+s_{\alpha}^{2}s_{2\beta}^{2}(c_{\theta}^{2}+s_{\theta}^{2}c_{\alpha}^{2})}{4\,c_{\theta}^{2}c_{\alpha}^{2}+s_{\alpha}^{2}s_{2\beta}^{2}(c_{\theta}^{2}+s_{\theta}^{2}c_{\alpha}^{2})}\dd\psi,\\
A_{3} & =\left(1-\frac{4\,s_{\alpha}^{2}s_{2\beta}^{2}s_{\theta}^{2}c_{\alpha}^{2}}{4\,c_{\theta}^{2}c_{\alpha}^{2}+s_{\alpha}^{2}s_{2\beta}^{2}(c_{\theta}^{2}+s_{\theta}^{2}c_{\alpha}^{2})}\right)\dd\psi.
\end{split}
\end{equation}

\subsubsection*{\texorpdfstring{Q$^{1,1,1}$}{Q111}\label{par:Q111_def}}

The solution for Q$^{1,1,1}$ is given in~\cite{AV05,GLMW05}. The angles parametrising the three-torus are
\begin{equation}
\varphi_{1}=\phi_{1},\quad\varphi_{2}=\phi_{2},\quad\varphi_{3}=\phi_{3}.
\end{equation}
The quantities $\Sigma$, $\vol_{4}$ and the $A_{a}$ are
\begin{equation}
\begin{split}\Sigma & =\frac{2\,c_{\theta_{3}}^{2}s_{\theta_{1}}^{2}s_{\theta_{2}}^{2}+(2-c_{2\theta_{1}}-c_{2\theta_{2}})s_{\theta_{3}}^{2}}{2048},\\
\vol_{4} & =8^{-3/2}H^{-1/2}s_{\theta_{1}}s_{\theta_{2}}s_{\theta_{3}}\,\dd\theta_{1}\wedge\dd\theta_{2}\wedge\dd\theta_{3}\wedge\dd\psi,\\
A_{1} & =\frac{8\,c_{\theta_{1}}s_{\theta_{2}}^{2}s_{\theta_{3}}^{2}}{H}\dd\psi,\\
A_{2} & =\left(\frac{2-c_{2\theta_{1}}-c_{2\theta_{2}}}{2\,s_{\theta_{1}}^{2}c_{\theta_{2}}}+\frac{s_{\theta_{2}}^{2}c_{\theta_{3}}^{2}}{c_{\theta_{2}}s_{\theta_{3}}^{2}}\right)^{-1}\dd\psi,\\
A_{3} & =\frac{8\,c_{\theta_{3}}s_{\theta_{1}}^{2}s_{\theta_{2}}^{2}}{H}\dd\psi.
\end{split}
\end{equation}
The function $H$ that appears here is given by
\begin{equation}
H=5-3\,c_{2\theta_{3}}+c_{2\theta_{1}}(-3+c_{2\theta_{2}}+c_{2\theta_{3}})+c_{2\theta_{2}}(-3+2\,c_{\theta_{1}}^{2}c_{2\theta_{3}}).
\end{equation}

\subsubsection*{\texorpdfstring{M$^{1,1,1}$}{M111}\label{par:M111_def}}

The solution for M$^{1,1,1}$ is given in~\cite{AV05,GLMW05}. The angles parametrising the three-torus are
\begin{equation}
\varphi_{1}=\tilde{\phi},\quad\varphi_{2}=\phi,\quad\varphi_{3}=\psi.
\end{equation}
The quantities $\Sigma$, $\vol_{4}$ and the $A_{a}$ are\footnote{Note that this is not the same deformation as \cite{AV05}. As pointed out in \cite{GLMW05}, the $\text{T}^{3}$ action must commute with the Killing spinors to preserve supersymmetry. A simple way to show this is to check that both $\omega$ and $\Omega$ are invariant – we have done this for our choice of angles.}

\begin{equation}
\begin{split}\Sigma & =\tfrac{3}{262144}h\sin^{2}\mu,\\
\vol_{4} & =-\tfrac{3\sqrt{3}}{16}h^{-1/2}\cos\mu\sin\theta\sin\tilde{\theta}\sin^{2}\mu\,\dd\mu\wedge\dd\tilde{\theta}\wedge\dd\theta\wedge\dd\tau,\\
A_{1} & =-64\,h^{-1}\cos\tilde{\theta}\cos^{2}\mu\sin^{2}\theta\,\dd\tau,\\
A_{2} & =24\,h^{-1}\cos\theta\sin^{2}\tilde{\theta}\sin^{2}2\mu\,\dd\tau,\\
A_{3} & =8\,h^{-1}\sin^{2}\theta(3+5\cos2\mu+2\cos2\tilde{\theta}\sin^{2}\mu)\dd\tau.
\end{split}
\end{equation}
The function $h$ is
\begin{equation}
\begin{split}h & =8\sin^{2}\theta\cos2\mu\,(\cos2\tilde{\theta}+7)-6(\cos2\theta+3)\sin^{2}\tilde{\theta}\cos4\mu\\
 & \eqspace+\cos2\theta\,(\cos2\tilde{\theta}-33)-13\cos2\tilde{\theta}+45.
\end{split}
\end{equation}


\begin{thebibliography}{66}
	
	
	\bibitem{Maldacena99}
	J.~Maldacena, ``The large {N} limit of superconformal field theories and
	supergravity'', \href{http://dx.doi.org/10.1023/A:1026654312961}{{\em Int. J.
			Theor. Phys.} {\bfseries 38} (1999)1113--1133},
	\href{http://arxiv.org/abs/hep-th/9711200}{{\ttfamily arXiv:hep-th/9711200}}.
	
	\bibitem{Bashmakov:2017rko}
	V.~Bashmakov, M.~Bertolini, and H.~Raj, ``On non-supersymmetric conformal
	manifolds: field theory and holography'',
	\href{http://dx.doi.org/10.1007/JHEP11(2017)167}{{\em JHEP} {\bfseries 11}
		(2017)167},
	\href{http://arxiv.org/abs/1709.01749}{{\ttfamily arXiv:1709.01749 [hep-th]}}.
	
	\bibitem{Cordova:2016xhm}
	C.~Cordova, T.~T. Dumitrescu, and K.~Intriligator, ``Deformations of
	superconformal theories'',
	\href{http://dx.doi.org/10.1007/JHEP11(2016)135}{{\em JHEP} {\bfseries 11}
		(2016)135},
	\href{http://arxiv.org/abs/1602.01217}{{\ttfamily arXiv:1602.01217 [hep-th]}}.
	
	\bibitem{Strassler:1998iz}
	M.~J. Strassler, ``On renormalization group flows and exactly marginal
	operators in three dimensions'',
	\href{http://arxiv.org/abs/hep-th/9810223}{{\ttfamily arXiv:hep-th/9810223
			[hep-th]}}.
	
	\bibitem{Tachikawa06}
	Y.~Tachikawa, ``Five-dimensional supergravity dual of $a$-maximization'',
	\href{http://dx.doi.org/10.1016/j.nuclphysb.2005.11.010}{{\em Nucl. Phys.}
		{\bfseries B733} (2006)188--203},
	\href{http://arxiv.org/abs/hep-th/0507057}{{\ttfamily arXiv:hep-th/0507057}}.
	
	\bibitem{Gaiotto:2007qi}
	D.~Gaiotto and X.~Yin, ``Notes on superconformal {C}hern--{S}imons-matter
	theories'', \href{http://dx.doi.org/10.1088/1126-6708/2007/08/056}{{\em JHEP}
		{\bfseries 08} (2007)056},
	\href{http://arxiv.org/abs/0704.3740}{{\ttfamily arXiv:0704.3740 [hep-th]}}.
	
	\bibitem{MS08}
	D.~Martelli and J.~Sparks, ``Notes on toric {S}asaki--{E}instein
	seven-manifolds and {AdS}(4) / {CFT}(3)'',
	\href{http://dx.doi.org/10.1088/1126-6708/2008/11/016}{{\em JHEP} {\bfseries
			11} (2008)016},
	\href{http://arxiv.org/abs/0808.0904}{{\ttfamily arXiv:0808.0904 [hep-th]}}.
	
	\bibitem{Asnin:2009xx}
	V.~Asnin, ``On metric geometry of conformal moduli spaces of four-dimensional
	superconformal theories'',
	\href{http://dx.doi.org/10.1007/JHEP09(2010)012}{{\em JHEP} {\bfseries 09}
		(2010)012},
	\href{http://arxiv.org/abs/0912.2529}{{\ttfamily arXiv:0912.2529 [hep-th]}}.
	
	\bibitem{Akerblom:2009gx}
	N.~Akerblom, C.~Saemann, and M.~Wolf, ``Marginal deformations and 3-algebra
	structures'', \href{http://dx.doi.org/10.1016/j.nuclphysb.2009.08.012}{{\em
			Nucl. Phys.} {\bfseries B826} (2010)456--489},
	\href{http://arxiv.org/abs/0906.1705}{{\ttfamily arXiv:0906.1705 [hep-th]}}.
	
	\bibitem{BPS10}
	M.~Bianchi, S.~Penati, and M.~Siani, ``Infrared stability of {ABJ}-like
	theories'', \href{http://dx.doi.org/10.1007/JHEP01(2010)080}{{\em JHEP}
		{\bfseries 01} (2010)080},
	\href{http://arxiv.org/abs/0910.5200}{{\ttfamily arXiv:0910.5200 [hep-th]}}.
	
	\bibitem{BPS10b}
	M.~Bianchi, S.~Penati, and M.~Siani, ``Infrared stability of ${N} = 2$
	{C}hern--{S}imons matter theories'',
	\href{http://dx.doi.org/10.1007/JHEP05(2010)106}{{\em JHEP} {\bfseries 05}
		(2010)106},
	\href{http://arxiv.org/abs/0912.4282}{{\ttfamily arXiv:0912.4282 [hep-th]}}.
	
	\bibitem{CY10}
	C.-M. Chang and X.~Yin, ``Families of conformal fixed points of $n=2$
	{C}hern--{S}imons-matter theories'',
	\href{http://dx.doi.org/10.1007/JHEP05(2010)108}{{\em JHEP} {\bfseries 05}
		(2010)108},
	\href{http://arxiv.org/abs/1002.0568}{{\ttfamily arXiv:1002.0568 [hep-th]}}.
	
	\bibitem{BP11}
	M.~Bianchi and S.~Penati, ``The conformal manifold of {C}hern--{S}imons matter
	theories'', \href{http://dx.doi.org/10.1007/JHEP01(2011)047}{{\em JHEP}
		{\bfseries 01} (2011)047}, \href{http://arxiv.org/abs/1009.6223}{{\ttfamily
			arXiv:1009.6223 [hep-th]}}.
	
	\bibitem{ALMTW14}
	S.~de~Alwis, J.~Louis, L.~McAllister, H.~Triendl, and A.~Westphal, ``Moduli
	spaces in {AdS}$_{4}$ supergravity'',
	\href{http://dx.doi.org/10.1007/JHEP05(2014)102}{{\em JHEP} {\bfseries 05}
		(2014)102}, \href{http://arxiv.org/abs/1312.5659}{{\ttfamily arXiv:1312.5659
			[hep-th]}}.
	
	\bibitem{Zamolodchikov:1986gt}
	A.~B. Zamolodchikov, ``Irreversibility of the flux of the renormalization group
	in a 2{D} field theory'', {\em JETP Lett.} {\bfseries 43} (1986)730--732.
	[Pisma Zh. Eksp. Teor. Fiz.43,565(1986)].
	
	\bibitem{LS95}
	R.~Leigh and M.~Strassler, ``Exactly marginal operators and duality in four
	dimensional $\mathcal{N} = 1$ supersymmetric gauge theory'',
	\href{http://dx.doi.org/10.1016/0550-3213(95)00261-P}{{\em Nucl. Phys.}
		{\bfseries B447} (1995)95--136},
	\href{http://arxiv.org/abs/hep-th/9503121}{{\ttfamily arXiv:hep-th/9503121}}.
	
	\bibitem{Kol02}
	B.~Kol, ``On conformal deformations'',
	\href{http://dx.doi.org/10.1088/1126-6708/2002/09/046}{{\em JHEP} {\bfseries
			09} (2002)046}, \href{http://arxiv.org/abs/hep-th/0205141}{{\ttfamily
			arXiv:hep-th/0205141}}.
	
	\bibitem{Kol10}
	B.~Kol, ``On conformal deformations {II}'',
	\href{http://arxiv.org/abs/1005.4408}{{\ttfamily arXiv:1005.4408 [hep-th]}}.
	
	\bibitem{GKSTW10}
	D.~Green, Z.~Komargodski, N.~Seiberg, Y.~Tachikawa, and B.~Wecht, ``Exactly
	marginal deformations and global symmetries'',
	\href{http://dx.doi.org/10.1007/jhep06(2010)106}{{\em JHEP} {\bfseries 06}
		(2010)106}, \href{http://arxiv.org/abs/1005.3546}{{\ttfamily arXiv:1005.3546
			[hep-th]}}.
	
	\bibitem{Baggio:2017mas}
	M.~Baggio, N.~Bobev, S.~M. Chester, E.~Lauria, and S.~S. Pufu, ``Decoding a
	three-dimensional conformal manifold'',
	\href{http://dx.doi.org/10.1007/JHEP02(2018)062}{{\em JHEP} {\bfseries 02}
		(2018)062},
	\href{http://arxiv.org/abs/1712.02698}{{\ttfamily arXiv:1712.02698 [hep-th]}}.
	
	\bibitem{AKY02}
	O.~Aharony, B.~Kol, and S.~Yankielowicz, ``On exactly marginal deformations of
	$\mathcal{N}=4$ {SYM} and type {IIB} supergravity on
	{AdS}$_{5}\times${S}$^{5}$'',
	\href{http://dx.doi.org/10.1088/1126-6708/2002/06/039}{{\em JHEP} {\bfseries
			06} (2002)039}, \href{http://arxiv.org/abs/hep-th/0205090}{{\ttfamily
			arXiv:hep-th/0205090}}.
	
	\bibitem{GP01}
	M.~Gra\~{n}a and J.~Polchinski, ``Supersymmetric three-form flux perturbations
	on {AdS}$_5$'', \href{http://dx.doi.org/10.1103/PhysRevD.63.026001}{{\em
			Phys. Rev.} {\bfseries D63} (2001)026001},
	\href{http://arxiv.org/abs/hep-th/0009211}{{\ttfamily arXiv:hep-th/0009211}}.
	
	\bibitem{LM05}
	O.~Lunin and J.~Maldacena, ``Deforming field theories with {U}(1)$\times${U}(1)
	global symmetry and their gravity duals'',
	\href{http://dx.doi.org/10.1088/1126-6708/2005/05/033}{{\em JHEP} {\bfseries
			05} (2005)033}, \href{http://arxiv.org/abs/hep-th/0502086}{{\ttfamily
			arXiv:hep-th/0502086}}.
	
	\bibitem{BH05}
	S.~Benvenuti and A.~Hanany, ``Conformal manifolds for the conifold and other
	toric field theories'',
	\href{http://dx.doi.org/10.1088/1126-6708/2005/08/024}{{\em JHEP} {\bfseries
			08} (2005)024}, \href{http://arxiv.org/abs/hep-th/0502043}{{\ttfamily
			arXiv:hep-th/0502043}}.
	
	\bibitem{BFMMPZ08}
	A.~Butti, D.~Forcella, L.~Martucci, R.~Minasian, M.~Petrini, and A.~Zaffaroni,
	``On the geometry and the moduli space of beta-deformed quiver gauge
	theories'', \href{http://dx.doi.org/10.1088/1126-6708/2008/07/053}{{\em JHEP}
		{\bfseries 07} (2008)053}, \href{http://arxiv.org/abs/0712.1215}{{\ttfamily
			arXiv:0712.1215 [hep-th]}}.
	
	\bibitem{GLMW05}
	J.~Gauntlett, S.~Lee, T.~Mateos, and D.~Waldram, ``Marginal deformations of
	field theories with {AdS}$_4$ duals'',
	\href{http://dx.doi.org/10.1088/1126-6708/2005/08/030}{{\em JHEP} {\bfseries
			08} (2005)030},
	\href{http://arxiv.org/abs/hep-th/0505207}{{\ttfamily arXiv:hep-th/0505207}}.
	
	\bibitem{AV05}
	C.~Ahn and J.~Vazquez-Poritz, ``Marginal deformations with {U}(1)$^3$ global
	symmetry'', \href{http://dx.doi.org/10.1088/1126-6708/2005/07/032}{{\em JHEP}
		{\bfseries 07} (2005)032},
	\href{http://arxiv.org/abs/hep-th/0505168}{{\ttfamily arXiv:hep-th/0505168}}.
	
	\bibitem{ABJM08}
	O.~Aharony, O.~Bergman, D.~Jafferis, and J.~Maldacena, ``$\mathcal{N}=6$
	superconformal {C}hern--{S}imons-matter theories, {M}2-branes and their
	gravity duals'', \href{http://dx.doi.org/10.1088/1126-6708/2008/10/091}{{\em
			JHEP} {\bfseries 10} (2008)091},
	\href{http://arxiv.org/abs/0806.1218}{{\ttfamily arXiv:0806.1218 [hep-th]}}.
	
	\bibitem{CSW11}
	A.~Coimbra, C.~Strickland-Constable, and D.~Waldram, ``{E}$_{d(d)} \times
	\mathbb{R}^+$ generalised geometry, connections and {M} theory'',
	\href{http://dx.doi.org/10.1007/JHEP02(2014)054}{{\em JHEP} {\bfseries 02}
		(2014)054}, \href{http://arxiv.org/abs/1112.3989}{{\ttfamily arXiv:1112.3989
			[hep-th]}}.
	
	\bibitem{CSW14}
	A.~Coimbra, C.~Strickland-Constable, and D.~Waldram, ``Supergravity as
	generalised geometry {II}: {E}$_{d(d)} \times \mathbb{R}^+$ and {M} theory'',
	\href{http://dx.doi.org/10.1007/JHEP02(2014)054}{{\em JHEP} {\bfseries 03}
		(2014)019}, \href{http://arxiv.org/abs/1212.1586}{{\ttfamily arXiv:1212.1586
			[hep-th]}}.
	
	\bibitem{PW08}
	P.~Pacheco and D.~Waldram, ``{M}-theory, exceptional generalised geometry and
	superpotentials'',
	\href{http://dx.doi.org/10.1088/1126-6708/2008/09/123}{{\em JHEP} {\bfseries
			09} (2008)123}, \href{http://arxiv.org/abs/0804.1362}{{\ttfamily
			arXiv:0804.1362 [hep-th]}}.
	
	\bibitem{GLSW09}
	M.~Gra\~{n}a, J.~Louis, A.~Sim, and D.~Waldram, ``{E}$_{7(7)}$ formulation of
	$\mathcal{N}=2$ backgrounds'',
	\href{http://dx.doi.org/10.1088/1126-6708/2009/07/104}{{\em JHEP} {\bfseries
			07} (2009)104}, \href{http://arxiv.org/abs/0904.2333}{{\ttfamily
			arXiv:0904.2333 [hep-th]}}.
	
	\bibitem{AGGPW16}
	A.~Ashmore, M.~Gabella, M.~Graña, M.~Petrini, and D.~Waldram, ``Exactly
	marginal deformations from exceptional generalised geometry'',
	\href{http://dx.doi.org/10.1007/JHEP01(2017)124}{{\em JHEP} {\bfseries 01}
		(2017)124},
	\href{http://arxiv.org/abs/1605.05730}{{\ttfamily arXiv:1605.05730 [hep-th]}}.
	
	\bibitem{AW15b}
	A.~Ashmore, M.~Petrini, and D.~Waldram, ``The exceptional generalised geometry
	of supersymmetric {AdS} flux backgrounds'',
	\href{http://dx.doi.org/10.1007/JHEP12(2016)146}{{\em JHEP} {\bfseries 12}
		(2016)146},
	\href{http://arxiv.org/abs/1602.02158}{{\ttfamily arXiv:1602.02158 [hep-th]}}.
	
	\bibitem{MPZ06}
	R.~Minasian, M.~Petrini, and A.~Zaffaroni, ``Gravity duals to deformed {SYM}
	theories and generalized complex geometry'',
	\href{http://dx.doi.org/10.1088/1126-6708/2006/12/055}{{\em JHEP} {\bfseries
			12} (2006)055}, \href{http://arxiv.org/abs/hep-th/0606257}{{\ttfamily
			arXiv:hep-th/0606257}}.
	
	\bibitem{BKO+18}
	I.~Bakhmatov, .~Kelekci, E.~Ó~Colgáin, and M.~M. Sheikh-Jabbari, ``Classical
	{Y}ang--{B}axter equation from supergravity'',
	\href{http://dx.doi.org/10.1103/PhysRevD.98.021901}{{\em Phys. Rev.}
		{\bfseries D98} 2, (2018)021901},
	\href{http://arxiv.org/abs/1710.06784}{{\ttfamily arXiv:1710.06784 [hep-th]}}.
	
	\bibitem{BCS+18}
	I.~Bakhmatov, E.~Colgáin, M.~M. Sheikh-Jabbari, and H.~Yavartanoo,
	``{Y}ang--{B}axter deformations beyond coset spaces (a slick way to do
	{TsT})'', \href{http://dx.doi.org/10.1007/JHEP06(2018)161}{{\em JHEP}
		{\bfseries 06} (2018)161},
	\href{http://arxiv.org/abs/1803.07498}{{\ttfamily arXiv:1803.07498 [hep-th]}}.
	
	\bibitem{SW99}
	N.~Seiberg and E.~Witten, ``{String theory and noncommutative geometry}'',
	\href{http://dx.doi.org/10.1088/1126-6708/1999/09/032}{{\em JHEP} {\bfseries
			09} (1999)032},
	\href{http://arxiv.org/abs/hep-th/9908142}{{\ttfamily arXiv:hep-th/9908142
			[hep-th]}}.
	
	\bibitem{AFHS99}
	B.~Acharya, J.~Figueroa-O'Farrill, C.~Hull, and B.~Spence, ``Branes at conical
	singularities and holography'', {\em Adv. Theor. Math. Phys.} {\bfseries 2}
	(1999)1249--1286,
	\href{http://arxiv.org/abs/hep-th/9808014}{{\ttfamily arXiv:hep-th/9808014}}.
	
	\bibitem{GKVW09}
	J.~Gauntlett, S.~Kim, O.~Varela, and D.~Waldram, ``Consistent supersymmetric
	{K}aluza-{K}lein truncations with massive modes'',
	\href{http://dx.doi.org/10.1088/1126-6708/2009/04/102}{{\em JHEP} {\bfseries
			04} (2009)102}, \href{http://arxiv.org/abs/0901.0676}{{\ttfamily
			arXiv:0901.0676 [hep-th]}}.
	
	\bibitem{AW15}
	A.~Ashmore and D.~Waldram, ``Exceptional {C}alabi--{Y}au spaces: the geometry
	of $\mathcal{N}=2$ backgrounds with flux'',
	\href{http://dx.doi.org/10.1002/prop.201600109}{{\em Fortsch. Phys.}
		{\bfseries 65} 1, (2017)1600109},
	\href{http://arxiv.org/abs/1510.00022}{{\ttfamily arXiv:1510.00022 [hep-th]}}.
	
	\bibitem{BCL06}
	K.~Behrndt, M.~Cveti\v{c}, and T.~Liu, ``Classification of supersymmetric flux
	vacua in {M}-theory'',
	\href{http://dx.doi.org/10.1016/j.nuclphysb.2006.04.018}{{\em Nucl. Phys.}
		{\bfseries B749} (2006)25--68},
	\href{http://arxiv.org/abs/hep-th/0512032}{{\ttfamily arXiv:hep-th/0512032}}.
	
	\bibitem{DP04}
	G.~Dall'Agata and N.~Prezas, ``$\mathcal{N}=1$ geometries for {M} theory and
	type {IIA} strings with fluxes'',
	\href{http://dx.doi.org/10.1103/PhysRevD.69.066004}{{\em Phys. Rev.}
		{\bfseries D69} (2004)066004},
	\href{http://arxiv.org/abs/hep-th/0311146}{{\ttfamily arXiv:hep-th/0311146}}.
	
	\bibitem{Hitchin02}
	N.~Hitchin, ``Generalized {C}alabi--{Y}au manifolds'',
	\href{http://dx.doi.org/10.1093/qjmath/54.3.281}{{\em Quart. J. Math.}
		{\bfseries 54} (2003)281--308},
	\href{http://arxiv.org/abs/math/0209099}{{\ttfamily arXiv:math/0209099
			[math-dg]}}.
	
	\bibitem{Gualtieri04}
	M.~Gualtieri, ``Generalized complex geometry'',
	\href{http://arxiv.org/abs/math/0401221}{{\ttfamily arXiv:math/0401221
			[math.DG]}}.
	
	\bibitem{Hull07}
	C.~Hull, ``Generalised geometry for {M}-theory'',
	\href{http://dx.doi.org/10.1088/1126-6708/2007/07/079}{{\em JHEP} {\bfseries
			07} (2007)079}, \href{http://arxiv.org/abs/hep-th/0701203}{{\ttfamily
			arXiv:hep-th/0701203}}.
	
	\bibitem{Hitchin00}
	N.~Hitchin, ``The geometry of three-forms in six dimensions'', {\em J. Diff.
		Geom.} {\bfseries 55} (2000)547--576.
	
	\bibitem{BDD+09}
	L.~Borsten, D.~Dahanayake, M.~J. Duff, and W.~Rubens, ``Black holes admitting a
	{F}reudenthal dual'',
	\href{http://dx.doi.org/10.1103/PhysRevD.80.026003}{{\em Phys. Rev.}
		{\bfseries D80} (2009)026003},
	\href{http://arxiv.org/abs/0903.5517}{{\ttfamily arXiv:0903.5517 [hep-th]}}.
	
	\bibitem{LS12}
	P.~Levay and G.~Sarosi, ``Hitchin functionals are related to measures of
	entanglement'', \href{http://dx.doi.org/10.1103/PhysRevD.86.105038}{{\em
			Phys. Rev.} {\bfseries D86} (2012)105038},
	\href{http://arxiv.org/abs/1206.5066}{{\ttfamily arXiv:1206.5066 [hep-th]}}.
	
	\bibitem{BDF+13}
	L.~Borsten, M.~J. Duff, S.~Ferrara, and A.~Marrani, ``Freudenthal dual
	{L}agrangians'', \href{http://dx.doi.org/10.1088/0264-9381/30/23/235003}{{\em
			Class. Quant. Grav.} {\bfseries 30} (2013)235003},
	\href{http://arxiv.org/abs/1212.3254}{{\ttfamily arXiv:1212.3254 [hep-th]}}.
	
	\bibitem{EST13}
	R.~Eager, J.~Schmude, and Y.~Tachikawa, ``Superconformal indices,
	{Sasaki}--{Einstein} manifolds, and cyclic homologies'',
	\href{http://dx.doi.org/10.4310/ATMP.2014.v18.n1.a3}{{\em Adv. Theor. Math.
			Phys.} {\bfseries 18} 1, (2014)129--175},
	\href{http://arxiv.org/abs/1207.0573}{{\ttfamily arXiv:1207.0573 [hep-th]}}.
	
	\bibitem{ES15}
	R.~Eager and J.~Schmude, ``Superconformal indices and {M}2-branes'',
	\href{http://dx.doi.org/10.1007/JHEP12(2015)062}{{\em JHEP} {\bfseries 12}
		(2015)062},
	\href{http://arxiv.org/abs/1305.3547}{{\ttfamily arXiv:1305.3547 [hep-th]}}.
	
	\bibitem{CS16}
	A.~Coimbra and C.~Strickland-Constable, ``Supersymmetric backgrounds, the
	{K}illing superalgebra, and generalised special holonomy'',
	\href{http://dx.doi.org/10.1007/JHEP11(2016)063}{{\em JHEP} {\bfseries 11}
		(2016)063},
	\href{http://arxiv.org/abs/1606.09304}{{\ttfamily arXiv:1606.09304 [hep-th]}}.
	
	\bibitem{Benini:2009qs}
	F.~Benini, C.~Closset, and S.~Cremonesi, ``Chiral flavors and {M}2-branes at
	toric {CY}4 singularities'',
	\href{http://dx.doi.org/10.1007/JHEP02(2010)036}{{\em JHEP} {\bfseries 02}
		(2010)036},
	\href{http://arxiv.org/abs/0911.4127}{{\ttfamily arXiv:0911.4127 [hep-th]}}.
	
	\bibitem{Cheon:2011th}
	S.~Cheon, D.~Gang, S.~Kim, and J.~Park, ``Refined test of {AdS4/CFT3}
	correspondence for ${N}=2,3$ theories'',
	\href{http://dx.doi.org/10.1007/JHEP05(2011)027}{{\em JHEP} {\bfseries 05}
		(2011)027},
	\href{http://arxiv.org/abs/1102.4273}{{\ttfamily arXiv:1102.4273 [hep-th]}}.
	
	\bibitem{JKPS11}
	D.~Jafferis, I.~Klebanov, S.~Pufu, and B.~Safdi, ``Towards the {F}-theorem:
	${N}=2$ field theories on the three-sphere'',
	\href{http://dx.doi.org/10.1007/JHEP06(2011)102}{{\em JHEP} {\bfseries 06}
		(2011)102},
	\href{http://arxiv.org/abs/1103.1181}{{\ttfamily arXiv:1103.1181 [hep-th]}}.
	
	\bibitem{DFN84}
	R.~D'Auria, P.~Fre, and P.~van Nieuwenhuizen, ``${N}=2$ matter coupled
	supergravity from compactification on a coset ${G/H}$ possessing an
	additional {K}illing vector'',
	\href{http://dx.doi.org/10.1016/0370-2693(84)92018-5}{{\em Phys. Lett.}
		{\bfseries B136} (1984)347--353}.
	
	\bibitem{PP84}
	D.~Page and C.~Pope, ``Which compactifications of ${D}=11$ supergravity are
	stable?'',
	\href{http://dx.doi.org/10.1016/0370-2693(84)91275-9}{{\em Phys. Lett.}
		{\bfseries B144} (1984)346--350}.
	
	\bibitem{BRS10}
	N.~Benishti, D.~Rodr\'{i}guez-G\'{o}mez, and J.~Sparks, ``Baryonic symmetries
	and {M}5 branes in the {AdS}$_4$/{CFT}$_3$ correspondence'',
	\href{http://dx.doi.org/10.1007/JHEP07(2010)024}{{\em JHEP} {\bfseries 07}
		(2010)024}, \href{http://arxiv.org/abs/1004.2045}{{\ttfamily arXiv:1004.2045
			[hep-th]}}.
	
	\bibitem{FKR09}
	S.~Franco, I.~Klebanov, and D.~Rodr\'{i}guez-G\'{o}mez, ``M2-branes on
	orbifolds of the cone over ${Q}^{1,1,1}$'',
	\href{http://dx.doi.org/10.1088/1126-6708/2009/08/033}{{\em JHEP} {\bfseries
			08} (2009)033}, \href{http://arxiv.org/abs/0903.3231}{{\ttfamily
			arXiv:0903.3231 [hep-th]}}.
	
	\bibitem{PZ09}
	M.~Petrini and A.~Zaffaroni, ``${N}=2$ solutions of massive type {IIA} and
	their {C}hern--{S}imons duals'',
	\href{http://dx.doi.org/10.1088/1126-6708/2009/09/107}{{\em JHEP} {\bfseries
			09} (2009)107}, \href{http://arxiv.org/abs/0904.4915}{{\ttfamily
			arXiv:0904.4915 [hep-th]}}.
	
	\bibitem{FFGRTZZ00}
	D.~Fabbri, P.~Fr\'e, L.~Gualtieri, C.~Reina, A.~Tomasiello, A.~Zaffaroni, and
	A.~Zampa, ``3-{D} superconformal theories from {S}asakian seven manifolds:
	new nontrivial evidences for {AdS}(4) / {CFT}(3)'',
	\href{http://dx.doi.org/10.1016/S0550-3213(00)00098-5}{{\em Nucl. Phys.}
		{\bfseries B577} (2000)547--608},
	\href{http://arxiv.org/abs/hep-th/9907219}{{\ttfamily arXiv:hep-th/9907219
			[hep-th]}}.
	
	\bibitem{HT09}
	N.~Halmagyi and A.~Tomasiello, ``Generalized {K}\"{a}hler potentials from
	supergravity'', \href{http://dx.doi.org/10.1007/s00220-009-0881-6}{{\em
			Commun. Math. Phys.} {\bfseries 291} 1, (2009)1--30},
	\href{http://arxiv.org/abs/0708.1032}{{\ttfamily arXiv:0708.1032 [hep-th]}}.
	
	\bibitem{Dlamini:2016aaa}
	H.~Dlamini and K.~Zoubos, ``Integrable {H}opf twists, marginal deformations and
	generalised geometry'',
	\href{http://arxiv.org/abs/1602.08061}{{\ttfamily arXiv:1602.08061 [hep-th]}}.
	
	\bibitem{Mansson:2008xv}
	T.~Mansson and K.~Zoubos, ``Quantum symmetries and marginal deformations'',
	\href{http://dx.doi.org/10.1007/JHEP10(2010)043}{{\em JHEP} {\bfseries 10}
		(2010)043},
	\href{http://arxiv.org/abs/0811.3755}{{\ttfamily arXiv:0811.3755 [hep-th]}}.
	
	\bibitem{LSW14}
	K.~Lee, C.~Strickland‐Constable, and D.~Waldram, ``Spheres, generalised
	parallelisability and consistent truncations'',
	\href{http://dx.doi.org/10.1002/prop.201700048}{{\em Fortsch. Phys.}
		{\bfseries 65} 10-11, (2017)1700048},
	\href{http://arxiv.org/abs/1401.3360}{{\ttfamily arXiv:1401.3360 [hep-th]}}.
\end{thebibliography}
\end{document}